\documentclass{jfm}
\usepackage[utf8]{inputenc}
\usepackage{color}
\usepackage{ulem}
\newcommand\DEL[1]{\sout{}}     
\def\ADD#1{{\textcolor{black}{#1}}}    
\usepackage{graphicx}

\def\THI#1{{\textcolor{blue}{#1}}}    

\title[Added mass: a complex face of tidal conversion]
{Added mass: a complex face of tidal conversion}

\author[C. Brouzet,  E. V. Ermanyuk, M. Moulin, G. Pillet, T. Dauxois]
  {C. Brouzet\textsuperscript{1},  E. V. Ermanyuk\textsuperscript{1,2,3},  M. Moulin\textsuperscript{1}, G. Pillet\textsuperscript{1} and T.~Dauxois\textsuperscript{1}}

\affiliation{
  \textsuperscript{1}
 Univ Lyon, ENS de Lyon, Univ Claude Bernard, CNRS, Laboratoire de Physique, F-69342 Lyon, France\\
  [\affilskip]
  \textsuperscript{2}
  Lavrentyev Institute of Hydrodynamics, av. Lavrentyev 15, Novosibirsk 630090,
  Russia \\
  [\affilskip]
  \textsuperscript{3}
  Novosibirsk State University, Pirogova str. 2, Novosibirsk 630090,
  Russia \\
  }
\pubyear{???} \volume{???} \pagerange{???}
\date{\today}
\begin{document}

\maketitle

\begin{abstract}

This paper revisits the problem of tidal conversion at a ridge in a uniformly stratified fluid of limited \DEL{or infinite }depth \ADD{
 \ADD{using measurements of complex-valued added mass}. When the height of a sub-marine ridge is non negligible with respect to the depth of the water, the tidal conversion can be enhanced in the supercritical regime or reduced in the subcritical regime with respect to the large depth situation. Tidal conversion can even be null for some specific cases. Here, w}e study experimentally the \ADD{influence of finite depth on the} {added mass}\DEL{ and the radiated internal wave power } \ADD{coefficients} for \DEL{four}\ADD{three} different ridge\DEL{s, going beyond previous works} \ADD{shapes}. \DEL{We first show that the transition between subcritical and supercritical regimes of wave emission can be observed for a triangular ridge. Then, we}\ADD{We first} show that at low forcing frequency the tidal conversion is \ADD{weakly} enhanced by shallow depth for a semi-circular ridge\DEL{, a phenomenon predicted for a vertical sharp ridge}. In addition, \DEL{experiments carried out with}\ADD{added mass coefficients measured for} a vertical ridge show strong similarities with the ones \DEL{performed}\ADD{obtained} \DEL{with}\ADD{for} the semi-circular ridge. Nevertheless, the enhancement of the tidal conversion \ADD{at low forcing frequency} for the vertical ridge has not been observed, in contrast with \ADD{its supercritical shape}\DEL{the high-frequency behaviour of the added mass of the ridge which shows an overall good agreement with available theory}. Finally, we provide the experimental evidence of a lack of \DEL{radiated power}\ADD{tidal conversion} due to the specific shape of a ridge 
for certain depth and frequency tuning.

\end{abstract}

\section{Introduction}

In stratified oceans, the interaction of the tidal motion with the bottom topography continuously generates internal waves~\citep{Bell1975,Vlasenkoetal2005,GarrettKunze2007}. The global rate of energy conversion from barotropic to baroclinic tide is estimated to be around $1$~TW~\citep{Morozov1995,GarrettKunze2007}. 
\ADD{There exists an extensive literature on tidal conversion, which can be roughly splitted into two branches depending on the focus of the studies: i) oceanographic applications to various specific cases of bottom topography and stratification, or ii) structure of interior shear layers in internal wave beams generated by bodies of simple geometrical shape.}

The \DEL{``oceanographic"} literature \ADD{oriented toward oceanographic applications} describes a variety of mountain-shaped profiles~\citep{Bell1975, Baines1973,Baines1982, Craig1987, Balmforthetal2002, LlewellynSmithYoung2002, StLaurentetal2003, LlewellynSmithYoung2003, Khatiwala2003,BalmforthPeacock2009,Echeverrietal2009,EcheverriPeacock2010},
progressing from the flat subcritical topography in infinitely deep uniformly stratified fluid to more realistic cases with finite-slope bottom profiles, rotation, non-uniform stratification, limited depth, supercritical slopes, skewed mountain profiles and multiple ridges. In particular, recent studies on supercritical ridges in the ocean of limited depth\ADD{~\citep{StLaurentetal2003,LlewellynSmithYoung2003,Khatiwala2003,Petrelisetal2006,Echeverrietal2009,EcheverriPeacock2010,Rapakaetal2013}} have been motivated by applications to internal tides from the Hawaii ridge, which are well-documented in the field and satellite observations~\citep{EgbertRay2000,EgbertRay2001}. Regarding the effect of limited depth, which is the \DEL{main }focus of the present paper, two different patterns have been observed for isolated mountains. For a subcritical \DEL{elongated }topography \ADD{having a length scale of the same order as the horizontal wave length of the internal tide},~\cite{LlewellynSmithYoung2002} report the reduction of tidal conversion compared to the case of unlimited depth~\citep{Bell1975} by a factor ranging from~0 to~1. They attribute it to destructive interference of waves undergoing multiple reflections between the bottom and  free surface. For a sharp vertical ridge, an ultimate case of a supercritical topography, both~\cite{StLaurentetal2003} and~\cite{LlewellynSmithYoung2003} report the enhancement of tidal conversion compared to the case of infinite depth by a factor, increasing from 1 to infinity as the gap between the top of the ridge and the free surface decreases to zero. \ADD{The calculations of~\cite{Petrelisetal2006} for a triangular ridge in a fluid of limited depth have spanned the parameter space from the subcritical topography of~\cite{LlewellynSmithYoung2002} to the knife-edge ridge of~\cite{LlewellynSmithYoung2003} and have shown an enhancement of tidal conversion in the supercritical regime when the depth is reduced. 
} Obviously, for more complicated \ADD{and natural} topographies either of the two scenarios is possible, depending on the combination of \DEL{parameters}\ADD{the local conditions (stratification, frequency of the tides, geometry of the ridge, etc.). Moreover,}\DEL{. C} \ADD{c}onstructive and destructive interference of waves and wave focusing at a complex topography in a fluid of limited depth provides many exciting possibilities, including the physical curiosities like wave attractors between mountain ridges~\citep{Echeverrietal2011} and the lack of tidal conversion for ``well-tuned" topographic profiles~\citep{Petrelisetal2006,Maas2011}.

The \DEL{``fluid mechanics"} literature \ADD{focused on viscous effects in internal wave beams describes the development of interior shear layers due to}\DEL{has been mainly focused on the}
  oscillations of bodies \ADD{of simple geometry} in a uniformly stratified fluid of infinite extent. For horizontal oscillations with small amplitude (compared to the size of the body), this problem is equivalent to the problem of tidal conversion, with the virtual ``bottom" taken at the horizontal plane of symmetry of the body. The studied cases \DEL{of the body shapes} include elliptic cylinders in two dimensions~\citep{ApplebyCrighton1986,GorodtsovTeodorovich1986,Hurley1997,HurleyKeady1997} and ellipsoids and spheres in three 
 dimensions ~\citep{ApplebyCrighton1987,LaiChengMing1981,Kingetal2009,VEF2011}. Note that these shapes always have 
 parts with sub- and supercritical slopes, and that the limiting case of a very thin vertical elliptical cylinder~\citep{Hurley1997} is 
 equivalent to a steep ridge ~\citep{LlewellynSmithYoung2003} in infinite fluid. Considerable effort has been put into 
 regularization of the divergence at the characteristic lines tangent to the body surface by inclusion of viscous effects. The 
 theoretical solutions have been thoroughly verified experimentally by measurement of wave fields~\citep{Sutherlandetal1999,Sutherlandetal2000,SutherlandLinden2002,Zhangetal2007,VEF2011,EFV2011} and forces acting 
 on oscillating bodies~\citep{Ermanyuk2000,Ermanyuk2002}. The force measurements confirmed that, at the laboratory 
 scale, the radiated internal-wave power can be estimated from the ideal-fluid theory if the thickness of the boundary layer is 
 sufficiently small compared to the size of the body. The effect of limited depth on the radiated internal-wave power has been 
 studied numerically and experimentally for a circular cylinder~\citep{Sturova2001,ErmanyukGavrilov2002a} and 
 experimentally for a sphere~\citep{ErmanyukGavrilov2003}, demonstrating the reduction of the radiated wave power similar 
 to the subcritical case described in~\cite{LlewellynSmithYoung2002}, rather than the enhancement anticipated for the 
 supercritical case~\citep{LlewellynSmithYoung2003}. However, these studies were not fully conclusive because of the 
 limitations of the experimental techniques~\citep{ErmanyukGavrilov2002a} and also numerics~\citep{Sturova2001} at low 
 frequenc\DEL{y}\ADD{ies} of oscillations, \DEL{which is }of particular interest \DEL{in geophysical applications}\ADD{for supercritical situations}.

In this paper, we revisit the problem of tidal conversion \ADD{in a uniformly stratified fluid of limited depth} by considering an isolated bi-dimensional bottom topography\DEL{ in a uniformly stratified fluid of limited or infinite depth}, using the concepts of affine similitude and added mass~\citep{Ermanyuk2002,VEF2011}.
First, in section~\ref{theory_pendulum}, we introduce the theoretical preliminaries on added mass in homogeneous and stratified fluids. Then, in section~\ref{setup}, we present the experimental set-up and the data analysis. 
\DEL{Section~\ref{square} is dedicated to the case of a cylinder with a square-shaped cross section in a uniformly stratified fluid of infinite depth. 
The cross section has a vertically oriented diagonal. It serves as a model of a triangular mountain, with a sudden change from super- to subcritical slope at certain frequency.} 
Section~\ref{disk} describes the results obtained for a cylinder with a circular cross section\DEL{ in a fluid of finite depth}. This goes beyond the work performed by~\cite{ErmanyukGavrilov2002a} 
 and highlights a trend toward enhancement of the tidal conversion at small depth and frequency, which has been predicted by~\cite{LlewellynSmithYoung2003}
  for a vertical ridge. In section~\ref{vertical_plate}, we discuss experiments carried out with a vertical plate and show strong similarities between the \DEL{experimental results obtained}\ADD{added mass coefficients measured} with the vertical plate and with the cylinder having a circular cross section. Finally, in section~\ref{flattop}, a topography lacking tidal conversion for given frequencies~\citep{Petrelisetal2006,Maas2011} is tested. 
 Experiments with \DEL{this kind of}\ADD{such} topographies have only been reported recently~\citep{Pacietal2015} but the radiated wave power of such structures has never been measured yet.

\section{Theoretical preliminaries and set-up\label{theory_pendulum}}
\subsection{Added mass in homogeneous and stratified fluids}

When an object moves in an ideal fluid, it has to move the fluid around in order to pass through. As the fluid has a given density, a higher force is necessary to displace the object in the fluid than in the vacuum. 
Indeed, as both the object and the surrounding fluid have to be accelerated, the necessary force is equal to the acceleration of the 
object multiplied by the mass of the object 
and another mass $m_A$, due to the fluid~\citep{Lamb1932}. 
This mass is called the added mass and depends on the shape of the object in the fluid. It is generally described using a tensor. In naval architecture, the added mass plays an important role because it can easily \DEL{reach a large amount of}\ADD{be of the same order of magnitude as} the total mass of a ship or a submarine~\citep{Newman1977,Brennen1982}. 

Added mass can be used to investigate the tidal conversion in the oceans~\citep{VEF2011}, which are continuously stratified in density. In such a fluid, internal waves can propagate using buoyancy as a restoring force. 
One defines the buoyancy frequency $N=[(g/\bar{\rho})({\rm d}\rho/{\rm d}z)]^{1/2}$, where $\rho(z)$ is 
the density distribution over vertical coordinate~$z$, $\bar{\rho}$ a reference value and $g$  the gravity acceleration. \ADD{Note that, here, the  vertical axis~$z$ points downwards, as shown in figure~\ref{fig:schema_theory}.}
The dispersion relation of internal waves is $\omega/N=\pm\sin \theta$
where \DEL{the angle $\theta$ is the slope of the wave beam to the horizontal}\ADD{$\theta$ is the angle between the direction of propagation of the wave and the horizontal}, and $\omega$  the frequency of \DEL{oscillations}\ADD{the wave}. In the ocean, 
internal waves are created thanks to the tidal oscillations of the fluid upon the topographies. 
This can also be viewed as topographies oscillating in a stratified fluid. The two problems are totally equivalent for small amplitude oscillations.

Let us consider an object oscillating horizontally in a stratified fluid. With the dispersion relation, internal waves can only be emitted if the body oscillates at a frequency~$\omega$ smaller than $N$. This leads to two different behaviours of the body depending on the frequency of the oscillations. If $\omega>N$, no wave can be emitted, and the added mass is a real-valued function of $\omega$. But if $\omega<N$, waves are emitted by the body and the added mass becomes a complex-valued function of $\omega$. For unidirectional rectilinear oscillations of a body having three planes of symmetry, the added mass can be decomposed in two parts\ADD{~\citep{LaiChengMing1981,Ermanyuk2002}}
\begin{equation}
m_A=\mu(\omega)-\textrm{i}\frac{\lambda_{\textrm{w}}(\omega)}{\omega},\label{eq:def_added_mass}
\end{equation}
where \ADD{the real part $\mu$ corresponds physically to the force component oscillating in phase with the acceleration of the body (inertial force), and the imaginary part $\lambda_{\textrm{w}}(\omega)/{\omega}$ corresponds to the force component oscillating in phase with the velocity of the body (damping force).} $\mu$ is \ADD{therefore called} the inertial \ADD{mass}\DEL{coefficient, also called added mass by abuse of language,}
while $\lambda_{\textrm{w}}$ is the \ADD{wave} damping \DEL{coefficient}\ADD{rate, proportional to the radiated wave power as follows
\begin{equation}
     P(\omega)=\frac{1}{2}(A\omega)^2 \lambda_{\textrm{w}}(\omega),
    \label{eq:power}
   \end{equation}
$A$ being the amplitude of oscillations. The power  is directly related to the tidal conversion, with $U=A\omega$, the amplitude of tidal velocity.} Note that we assume an ideal \DEL{inviscid} fluid. In a real fluid, as in experiments, the motion of the object is affected by viscosity. Thus, in a uniformly stratified fluid when $\omega>N$, there is only a viscous damping while, when $\omega<N$, the damping is due to the combined effects of wave radiation and viscosity. Note that, in the latter case, the largest part of the energy dissipation is associated with the wave emission \ADD{(see for example~\cite{Ermanyuk2000,Ermanyuk2002})}.

\subsection{Affine similitude in a linearly stratified fluid}

\begin{figure}
\begin{center}
  \includegraphics[width=0.7\linewidth,clip=]{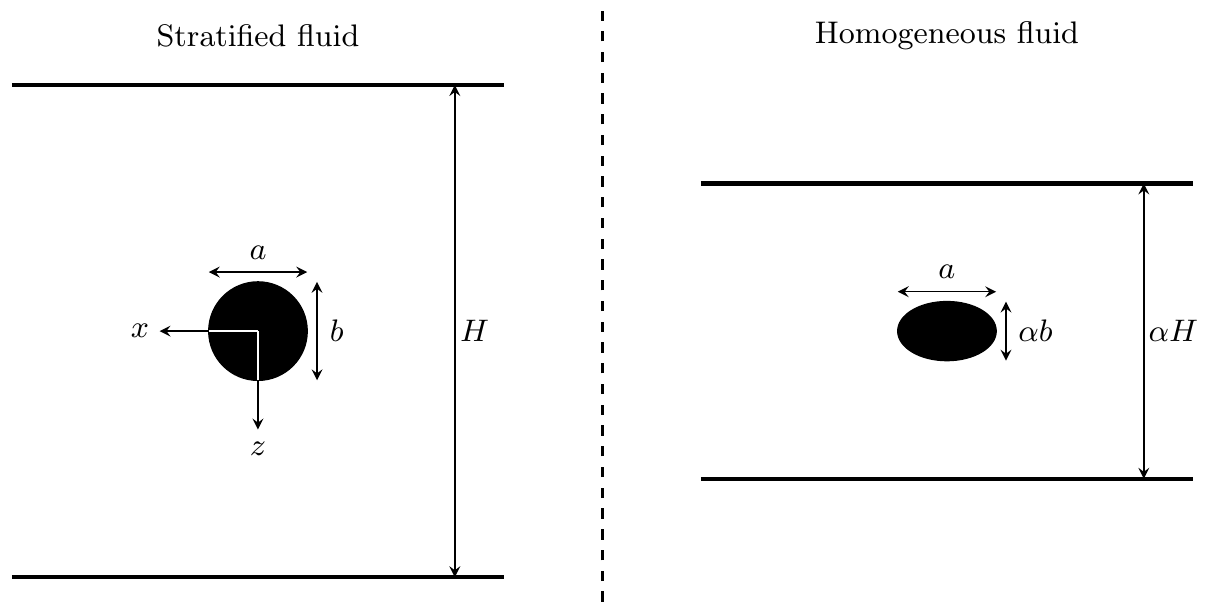}
    \caption{Geometries of the \DEL{real}\ADD{original} and fictitious problems. Left panel: two-dimensional object \ADD{oscillating horizontally at the non-dimensional frequency $\Omega$} in a linearly stratified fluid of depth $H$. The horizontal and vertical sizes of the object are named $a$ and $b$, respectively. Right panel: fictitious body \ADD{oscillating horizontally} in a homogeneous fluid after an affine transformation of the object in the left panel. The transformation changes the vertical scales by a factor $\alpha=(\Omega^2 -1)^{1/2}/ \Omega$, for $\Omega>1$. The horizontal and vertical sizes of the fictitious object are $a$ and $b_{\ast}=\alpha b$, respectively. The fluid has a depth of $\alpha H$. \label{fig:schema_theory}}
   \end{center}
\end{figure}

Let us consider a two-dimensional object submerged at depth $H/2$ in a channel of full depth $H$ filled with an ideal
uniformly
 stratified fluid, 
as sketched on the left panel in figure~\ref{fig:schema_theory}. We assume that the horizontal extent of the channel is infinite. The upper and lower boundaries of the channel are assumed to be rigid. A Cartesian coordinate system is introduced with the $x$-axis located at mid-depth of fluid and pointing to the left, and $z$-axis pointing downwards. The $y$-axis is horizontal and perpendicular to the $x$-axis. The horizontal and vertical sizes of the object are denoted $a$ and $b$, respectively. One defines the aspect ratio of the body $p=b/a$ and the \DEL{aspect ratio of the fluid}\ADD{non-dimensional vertical size of the body} $q=b/H$. We restrict our consideration to the case of horizontal harmonic oscillations of the object with frequency $\omega$ and amplitude $A$. The non-dimensional frequency is introduced as $\Omega=\omega /N$. Note that on the left panel in figure~\ref{fig:schema_theory}, the body is a circular cylinder but the theoretical preliminaries of this section are also valid for any other shapes. Below we consider experimentally the cases of cylinders with cross sections being \DEL{a square, }a disk, a line and a double-flattop. 
   In the oceanographic context, these geometries correspond to the \DEL{triangular, }semi-circular, \DEL{vertical}\ADD{knife-edge} and flattop ridges.

Let us consider first the problem where $\Omega >1$, i.e. where the added mass is a real quantity. The added mass coefficient $K$ of a body undergoing horizontal oscillations \ADD{at the dimensionless frequency $\Omega$} in uniformly stratified fluid are known~\citep{Ermanyuk2002} to be related with the added mass coefficient $K_{\ast}$ of a fictitious affinely similar body oscillating 
in a homogeneous fluid
\begin{equation}
K(\Omega)=K_{\ast}.\label{eq:k1k2}
\end{equation}
Here, the added mass coefficients are defined as $K=m_A/\rho_0 S$ and $K_{\ast}=m_{A\ast}/\rho_0 S_{\ast}$, where $\rho_0$ is the reference density at the depth corresponding to the centre of the body, $m_A$ and $m_{A\ast}$ are the added masses per unit length, $S$ and $S_{\ast}$ are the cross sections of the \DEL{initial}\ADD{original} and fictitious bodies, respectively. The fictitious \DEL{body}\ADD{problem} is obtained by compressing the \DEL{initial}\ADD{original} body \ADD{and channel} in the vertical direction $\alpha$ times, where $\alpha =(\Omega^2 -1)^{1/2}/ \Omega $. 
\DEL{The same relation applies in the fluid of infinite extent and in a horizontal \DEL{stripe}\ADD{channel} with height $H$ transformed into a \DEL{stripe}\ADD{channel} of depth $\alpha H$.}
This is shown in the right panel in figure~\ref{fig:schema_theory} where the fictitious body is an ellipse of major axis $a$ and minor axis $b_{\ast}=\alpha b$ \ADD{and where the fictitious channel has a height~$\alpha H$}. 
Consequently, the cross-section area $S=\pi a b/4$ of the \ADD{original} body is transformed in $S_{\ast}=\pi a b_{\ast}/4=\alpha S$ for the fictitious body. \ADD{Note that as the fictitious body oscillates in a homogeneous fluid, its added mass coefficient $K_{\ast}$ is independent of the frequency of oscillations. Therefore, the dependence on $\Omega$ for $K$ only comes from the dependence of $K_{\ast}$ on the compression coefficient $\alpha$. 
Moreover, when $\Omega$ becomes much greater than $1$, $\alpha$ tends to $1$. Thus, the original and fictitious problem geometries are the same for high oscillation frequencies. The added mass coefficient in the original problem for high frequency is supposed to tend to the one measured in the homogeneous case, with the same geometry.}

{In many problems, it is more convenient to normalize the added mass of the oscillating cylinder by the added mass of a flat plate of height $b$ so that $k=m_A/(\rho_0 \pi b^{2}/4)$ and $k_{\ast}=m_{A\ast}/(\rho_0 \pi b_{\ast}^{2}/4)$. This normalization is particularly suitable in geophysical fluid dynamics in view of the scaling used for the tidal conversion~\citep{LlewellynSmithYoung2003}. Obviously, with such a normalization, equation~(\ref{eq:k1k2}) should be replaced by}
\begin{equation}
k(\Omega)=k_{\ast}\alpha.
 \label {k1k2alpha}
\end{equation}

Equation~(\ref{eq:k1k2}) has been obtained by~\cite{Ermanyuk2002} by considering the integrals of pressure over the body surface and the control surface surrounding the body, and applying the Gauss theorem to the fluid volume located between these surfaces\ADD{, in spirit of~\cite{Newman1977}}. The control
surface can be a material surface that undergoes the same affine transformation as the body surface.  In this case, equation~(\ref{eq:k1k2}) remains valid since the conversion factors relating surface integrals over \DEL{initial}\ADD{original} and fictitious bodies as well as over \DEL{initial}\ADD{original} and transformed control surfaces are the same. However, \ADD{to extend the solution to the case} $\Omega <1 $, one should keep in mind that the control surface cannot be a closed one. There must be a possibility for radiation of internal wave energy. 

Now \ADD{we consider a body belonging to a certain family of shapes and oscillating in a
homogeneous fluid between two horizontal parallel rigid planes. L}\DEL{l}et us suppose that\DEL{, for a certain family of bodies oscillating in a
homogeneous fluid between two horizontal parallel rigid planes,}
we know the function, representing the dependence of the added
mass coefficients on non-dimensional geometrical parameters $p=b/a$ and $q=b/H$ \ADD{for this family of bodies}
\begin{equation}
K_{\ast}=F_{\ast} \left(p, q \right) {\hspace {10mm}} {\textrm{or}} {\hspace {10mm}} k_{\ast}=f_{\ast} \left(p, q \right).
\label {functions}
\end{equation}
Note that $q=b/H$ does not change under affine transformation. Then, for $\Omega>1$,
the added mass coefficient of a body in a uniformly stratified fluid at certain given values of $p$ and $q$ can
be found as follows
\begin{equation}
    K (\Omega ) = F_{\ast} \left(p\alpha, q \right) {\hspace {10mm}} {\textrm{or}} {\hspace {10mm}} k (\Omega ) = f_{\ast} \left(p\alpha, q \right)\alpha.
\label {rulell}
\end{equation}

As discussed in~\cite{Ermanyuk2002} in the context of~\cite{Hurley1997},
$K(\Omega )$ and $k(\Omega )$ at $\Omega<1$ 
can be obtained by the
analytic continuation in frequency. This can be done if the radiation condition
formulated in the causal sense does hold true, i.e. the
internal waves are radiated from the source to infinity and never
return back. For $\Omega<1$, the analytic continuation for $\alpha $ is $\textrm{i}\eta $, where $\eta = (1-\Omega ^{2})^{1/2}/\Omega $ is the
real-value parameter. Accordingly, equation~(\ref {rulell}) becomes
\begin{equation}
K (\Omega ) = F_{\ast} \left(p {\textrm i} \eta, q \right) {\hspace {10mm}} {\textrm{or}} {\hspace {10mm}} k (\Omega ) = f_{\ast} \left(p {\textrm i} \eta, q \right){\textrm i}\eta.
\label {rulhyp}
\end{equation}

Thus, knowing the added mass coefficient for a family of bodies in a homogeneous fluid, one can deduce the added mass coefficient of related bodies in a linearly stratified fluid, as a function of the frequency. This can be useful because added mass in a homogeneous fluid is well studied~\citep{Brennen1982,Korotkin2010} while the added mass in a stratified fluid has been investigated only recently.

When $\Omega<1$, the added mass is complex-valued, as defined by equation~(\ref{eq:def_added_mass}). 
One can introduce the \DEL{non-dimensional }inertial coefficient
\begin{eqnarray}
C^{\mu}=\textrm{Re} (k)=\textrm{Re} (K)\frac{4S}{\pi b^2}=\frac{4\mu}{\rho_0 \pi b^2},\label{eq:flat_plate_mu}
\end{eqnarray}
and the \ADD{wave} damping coefficient
\begin{eqnarray}
C^{\lambda}=\Omega~{\textrm{Im} (k)}=\Omega~{\textrm{Im} (K)}\frac{4S}{\pi b^2}=\frac{4\lambda_{\textrm{w}}}{\rho_0 \pi b^{2} N}.\label{eq:flat_plate_lambda}
\end{eqnarray}
\DEL{The damping coefficient is related to the radiated wave power as follows}
\DEL{where $A$ is the amplitude of oscillations. The power  is directly related to the tidal conversion, with $U=A\omega$, the amplitude of tidal velocity. T}\ADD{Using equation~(\ref{eq:power}), t}he dimensionless form for the radiated wave power is defined as
\begin{equation}
     P_{\textrm{w}}=\frac{1}{2}\Omega^2 C^{\lambda}(\Omega).
      \label{eq:power_coefficient}
   \end{equation}
\ADD{Note that in the case of an ideal and homogeneous fluid without free surface, one can define only the inertial coefficient $C^{\mu}$, the damping being identically zero.} \DEL{but not the wave damping coefficient $C^{\lambda}$. 
}

In the case of a fluid of infinite extent, 
equations~(\ref {rulell}) and~(\ref {rulhyp}) can be used to re-derive the formulas for hydrodynamic loads acting on an elliptic cylinder~\citep{Hurley1997}, and a vertically oscillating spheroid~\citep{LaiChengMing1981}. \DEL{New s}\ADD{S}olutions \ADD{for horizontally oscillating spheroids}\DEL{ can be also obtained(see }~\citep{Ermanyuk2002} and \ADD{squares}~\citep{ErmanyukGavrilov2002b} \ADD{can also be obtained}\DEL{)}. \ADD{However,} there are only few analytical solutions for the added mass of bodies oscillating in ideal homogeneous fluid of finite depth. The known results are limited to the cases of the vertical flat plate~\citep{LockwoodTaylor1930}, the elliptic cylinder~\citep{Clarke2001} and the rectangle~\citep{Gurevich1940,Newman1969}.

\section{Set-up and data processing\label{setup}}

\subsection{Experimental set-up}

\begin{figure}
\begin{center}
   \includegraphics[width=0.75\linewidth,clip=]{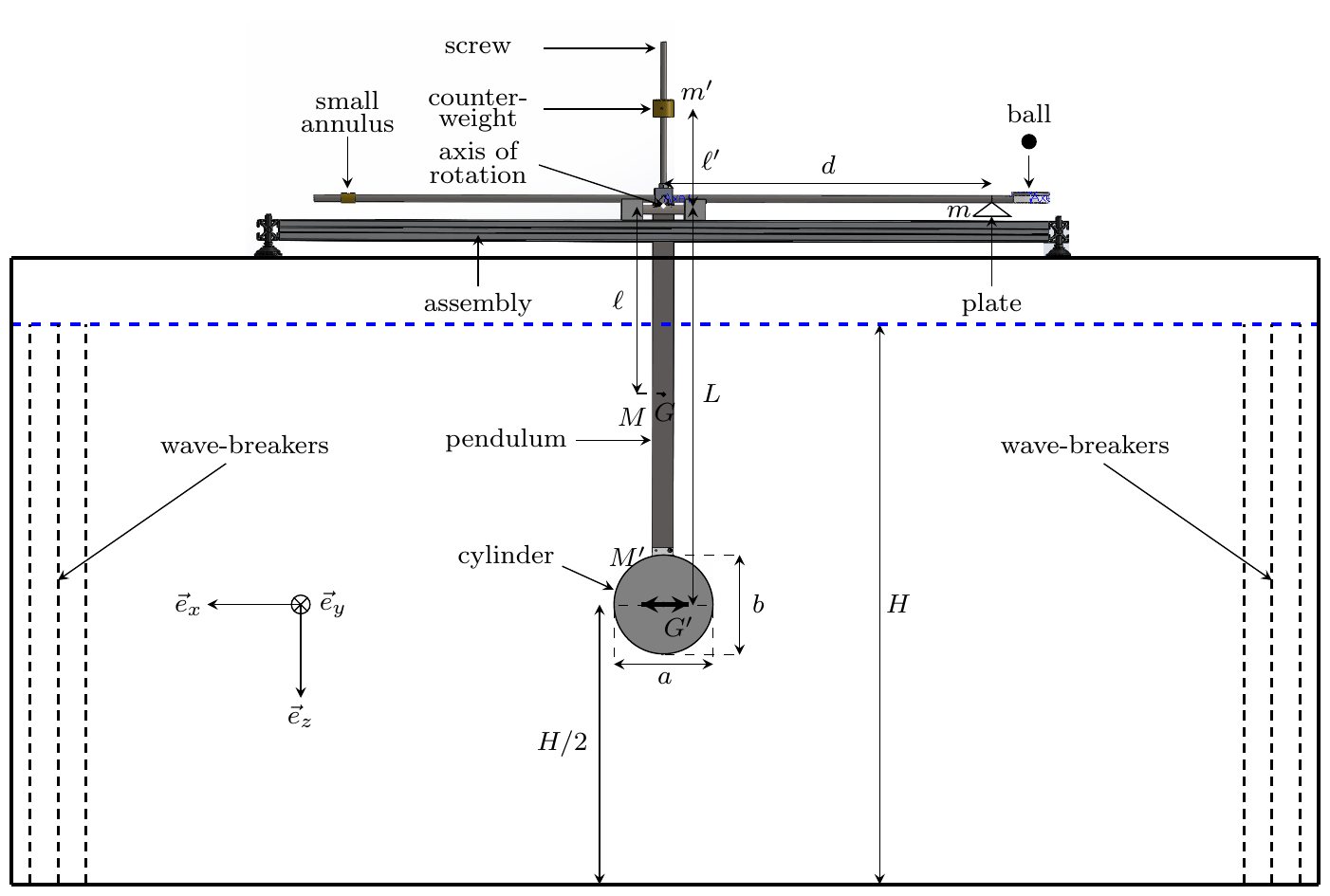}
    \caption{
    Schema of the pendulum set-up viewed from the side: arrows show the different elements of the set-up, relevant lengths are labelled and coordinates are shown. The surface of the water is the horizontal dashed blue line. \ADD{The body is moving horizontally along the $x$-axis, as depicted by the thick arrows in the center of the body.}\label{fig:setup_pendulum}}
   \end{center}
\end{figure}

The experimental set-up is sketched in figure~\ref{fig:setup_pendulum}. It is very similar to the one described by~\cite{Ermanyuk2000,Ermanyuk2002} and by~\cite{ErmanyukGavrilov2002a,ErmanyukGavrilov2003}. The pendulum has a cross shape, with a massive cylinder attached at the lower end. The cylinder has a center of gravity named $G'$ and a mass $M'$ and is invariant in the $y$-direction. It can have different cross section shapes, a \DEL{square}\ADD{disk} as in figure~\ref{fig:setup_pendulum} or {other shapes (see Table~\ref{tabular:shape_parameters})}. The cross section has a horizontal (respectively vertical) length scale noted $a$ (resp. $b$). 
The vertical arm opposite to the cylinder is a screw where \DEL{can lie }two different counter-weights of mass $m'=167$~g and $m'=704$~g \ADD{can be placed}. Changing the position of the counter-weight\ADD{s} with respect to the axis of rotation allows us to tune the characteristic frequency of the pendulum, defined as $\omega_c$. \ADD{The purpose of having two different counter-weights is to be able to cover a large range of frequencies for all the shapes studied in this article} 
The pendulum has also two long horizontal arms. At the end of the right one, there is a small horizontal circle, covered by a tensioned rubber membrane. A ball, initially hold by an electric magnet, can be dropped on the membrane to excite the pendulum. 
 As the period of the oscillations of the pendulum is of several seconds, if the membrane is sufficiently tight, the excitation of the pendulum is very close to an instantaneous impulse. Attached to the right arm, one also has a small plate
  which can contain a mass $m$ for the calibration procedure explained further, in section~\ref{equations_pendulum}. On the left horizontal arm, there is a small annulus which can be displaced horizontally  to adjust very precisely the horizontality of the pendulum before the beginning of each experiment. 
The center of mass of the pendulum \ADD{with the cylinder but} without the counter-weight is noted $G$, its mass $M$ and its moment of inertia is defined as $J$. Note that the mass of the pendulum alone, without the \DEL{object}\ADD{cylinder} and the counter-weight is $889$~g. All the relevant distances are defined in figure~\ref{fig:setup_pendulum}, with respect to the rotation axis. This axis is shown by a white spot in figure~\ref{fig:setup_pendulum}. $L$ and $\ell$ are given in Table~\ref{tabular:shape_parameters} for each cylinder. $d$ is equal to $40$~cm while $\ell'$ varies in the range $[5.1-21]$~cm. The different masses are also shown in figure~\ref{fig:setup_pendulum}. The coordinates are defined in figure~\ref{fig:setup_pendulum}, with the origin of the coordinate system taken at the center of mass of the cylinder, $G'$.

The lower end of the pendulum with the cylinder is immersed at mid-depth of a stably density stratified fluid of depth $H$, contained in a tank also sketched in figure~\ref{fig:setup_pendulum}. Two tanks were used to achieve different fluid depths. Experiments with $H=95$~cm were performed in a rectangular test tank of size $200 \times 17 \times 100 ~$cm$^{3}$ while the experiments with all the other fluid depths were conducted in a rectangular tank of size $160 \times 17 \times 42.5 ~$cm$^{3}$. The tank is filled with uniformly stratified fluid using the conventional double-bucket technique and using salt as a stratifying agent. The density profile is measured prior to experiments by a conductivity probe attached to a vertical traverse mechanism. The value of the buoyancy frequency $N$ is evaluated from the measured density profile. Note that the volume of the immersed streamlined part of the pendulum is less than $1\%$ of the volume of the cylinder. Thus, one can neglect its influence on the fluid-body interactions. The cylinder has a width $W'$ slightly smaller than $W$, the width of the tank, to avoid friction on the lateral walls of the tank. 
For experiments performed with a small depth, the frequency of the surface seiche modes \ADD{of the tank} can be close to the characteristic frequency of the pendulum $\omega_c$. Thus, it is possible to have an energy transfer from the pendulum to the surface seiche modes. A rigid lid placed on the free surface prevents the surface wave propagation and thus eliminates the energy transfers.

The pendulum is supported by an assembly which is fixed upon the tank. The rotation is made possible by two wedge-shape supports made of very strong steel, attached to the pendulum. Each wedge is in contact with a horizontal cylinder made of steel, oriented perpendicularly to the rib of the wedge and fixed on the assembly. 
With this arrangement, the friction at the contact points can be safely neglected. \ADD{The rotation of the pendulum is limited to very small angles so we can consider the motion of the cylinder as horizontal. This is shown by the two thick arrows in the center of the body in figure~\ref{fig:setup_pendulum}.}

The horizontal displacement $x$ of the cylinder is measured as a function of time using a laser and a position sensor, assuming the pendulum oscillates within very small angles. Experimentally, we took care to limit the forcing in order to have a maximal horizontal displacement of $1$~cm. The laser spot is reflected on a small mirror, fixed on the pendulum and located on the axis of rotation. These small details are not shown  in figure~\ref{fig:setup_pendulum}. As the pendulum oscillates, the mirror deflects the laser beam by an angle twice larger 
 than the angular displacement. The deviation is measured via an elongated position sensor, giving a voltage proportional to the deviation of the laser spot.   The deflection of the laser beam is recorded at a frequency of $50$~Hz during the experiment. Using the angular displacement and the distance $L$ between the rotation axis and the center of mass of the cylinder, one can get the horizontal displacement $x$.

To prevent the internal-wave reflection, \DEL{wave-breakers}\ADD{wave absorbers} are placed at each end  of the tank (see figure~\ref{fig:setup_pendulum}). They are composed of a network of five layers of grids with two different mesh sizes, $3$ and $1$~cm. 
 The space between the grids is around $5$~cm.

\subsection{Data acquisition procedure and impulse response function analysis\label{equations_pendulum}}

 Experiments have been performed, with four different cylinders and for a set of different fluid depths. For each cylinder, the moment of inertia $J$ and the distance $\ell$ between the rotation axis and the center of mass $G$ have been measured in air.  Table~\ref{tabular:shape_parameters} summarizes the different characteristics of the four cylinders used.

\begin{table}
\begin{center}
\begin{tabular}{l|c||c|c|c|c}
Section & $\S$ & App. \ref{square} & \ref{disk} & \ref{vertical_plate} & \ref{flattop}\\
 \hline
 \hline
Cross section shape & Unit & Square &  Circle  & Flat vertical plate & Flattop Hill \\
 \hline
 $a$ & cm & $14$  & $5$ & {$0.3$} & $20$\\
$b$ & cm & $14$ & $5$ & {$10$} & $8$\\
$H$ & cm & $95$ & $6-16$ & {$13-22$} & $12-95$\\
$H/b=1/q$ &  & $6.8$ & $1.2-3.2$ & {$1.3-2.2$} & $1.5-11.9$\\
$b/a=p$ &  & $1$ & $1$ & {$30$} & $0.4$\\
\hline
$W'$ & cm &  $17$  & $16$ & {$15.5$} & $16.4$\\
$\rho_c$ & g/cm$^3$ & $1.08$ & $1.43$ & {$2.86$} & $1.21$\\
$M$ & g & $2730$  & $1339$ & {$1022$} & $2833$ \\
$L$ & cm & $57.5$ & $56.7$ & {$55.5$} & $55.3$\\
$\ell$~(measured) & cm & $41.0\pm0.3$  & $22.3\pm0.3$ & {$19.2\pm0.4$} & $39.7\pm0.2$\\
$J$~(measured) & g$\cdot$m$^2$ & $655\pm4$  & $173\pm3$ & {$127\pm2$} & $639\pm5$\\
\end{tabular}
\caption{Parameters of the four cylinders used in \ADD{the different sections of} this paper. The shape indicates the cross section of the cylinder, in the $x-z$ plane. $a$ (respectively $b$) corresponds to the horizontal (resp. vertical) dimension of the cross section. \ADD{The fluid depth $H$ indicates the range of depths explored in this paper.} $W'$ is its width in the $y$-direction and $\rho_c$ its density. $M$ is the total mass of the pendulum and the cylinder, without the counter-weight. $L$ and $\ell$ are defined in figure~\ref{fig:setup_pendulum} while $J$ is the moment of inertia of the whole pendulum (without the counter-weight). Note that $J$ and $\ell$ are carefully measured in air. \DEL{The fluid depth $H$ indicates the range of fluid heights explored in this paper.} \ADD{The horizontal line in the middle of the table separates the parameters that determine the added mass (top) to the ones that are simple characteristics of the cylinders used (bottom).}}
\label{tabular:shape_parameters}
\end{center}
\end{table}

In a homogeneous or a stratified fluid, the equation of \DEL{forced }horizontal oscillations~$x$ of the centre of mass of the cylinder{ in the frequency domain} can be written~\citep{Cummins1962} as follows 
\begin{equation}
\left(\frac{J+m'\ell'^2+\mu(\omega) L^2}{L^2}\right)
{\ddot{x}}+\lambda(\omega) 
\,{\dot{x}} + \left(\frac{C-\ell' m' g}{L^2}\right) x = \THI{A \exp(i\omega t),}\label{eq:eau_oscillations}
\end{equation}
with $C$ being the constant containing the effects of the pendulum torque and of the buoyancy force. $\mu(\omega)$ is the \DEL{added}\ADD{inertial} mass and $\lambda(\omega)$ the damping \DEL{coefficient}\ADD{rate} of the fluid. \ADD{For a homogeneous fluid, this rate is defined as $\lambda_{\textrm{h}}$ and is only due to the viscosity of the fluid. For a stratified fluid, $\lambda$ is the sum of the wave damping rate $\lambda_{\textrm{w}}$, due to emission of internal waves, and the viscous damping rate $\lambda_{\textrm{h}}$.} The $C$ constant can be determined \ADD{during a static calibration:}\DEL{by loading} the plate \ADD{located} on the right arm of the pendulum \ADD{(see figure~\ref{fig:setup_pendulum}) is loaded with different masses $m$. This causes a deviation of the equilibrium position of the pendulum. Thus, the two first terms of the left hand side of equation~(\ref{eq:eau_oscillations}) are null (static calibration) while the right hand side is different from $0$ and depends on the masses $m$ and the geometry of the pendulum. Knowing everything except $C$, we can measure this constant}.

To determine $\mu$ and $\lambda$, we use \ADD{the classic concept} of impulse response function analysis 
(see e.g. \cite{Cummins1962}). The idea is to examine the response of \ADD{the pendulum}\DEL{equation~(\ref{eq:eau_oscillations})} to a forcing proportional to $\exp(i \omega t)$. The response function is defined by
\begin{equation}
R(\omega)=\int_0^{\infty}x(t')\exp(i\omega t')\textrm{d}t'.\label{eq:def_R}
\end{equation}
Using equations~(\ref{eq:eau_oscillations}) and~(\ref{eq:def_R}), one has the complex quantity
\begin{equation}
R(\omega)=\frac{1}{(C-\ell' m' g)+ i L^2 \lambda \omega-(J+m'\ell'^2+\mu L^2)\omega^2}.
\end{equation}
Denoting $|R(\omega)|$ its modulus and $\phi(\omega)$ its phase,
this leads to
\begin{eqnarray}
\mu&=&\frac{(C-\ell' m' g)}{\omega^2L^2}\left(1-\frac{|R(0)|}{|R(\omega)|}\cos \phi(\omega) \right)-\frac{J+m'\ell'^2}{L^2},\label{eq:mu}\\
\lambda&=&\frac{(C-\ell' m' g)}{L^2 \omega}\frac{|R(0)|}{|R(\omega)|}\sin \phi(\omega),\label{eq:lambda}
\end{eqnarray}
where $|R(0)|$ stands for the modulus of the impulse response function for $\omega=0$~rad/s. As the system is linear, the normalization of $|R(\omega)|$ by $|R(0)|$ allows us to use this Fourier transform approach at 
a small arbitrary  impulse. 
\DEL{In order to get the damping {coefficient} due to the wave emission $\lambda_{\textrm{w}}$, it is necessary to subtract to $\lambda$ the damping due to the viscosity of the water, $\lambda_{\textrm{h}}$, measured in a homogeneous fluid. The non-dimensional force and wave-power coefficients for a cylinder of unit length in direction $y$ are defined in equations~(\ref{eq:flat_plate_mu}),~(\ref{eq:flat_plate_lambda}) and~(\ref{eq:power_coefficient}).}
\ADD{For each position of the counter-weights, experiments are repeated three times and three signals $x(t)$ are recorded. Each measurement appears to be very reproducible. Then, the Fourier transform of each of the three signals is calculated to obtain three impulse response functions, defined in equation~(\ref{eq:def_R}). These three response functions are then averaged: the result is a complex function and its modulus and phase are computed. They are used to obtain the functions $\mu$ and $\lambda$ as a function of the frequency $\omega$.} The impulse response function analysis allows us to measure \DEL{the added mass}\ADD{$\mu$} and \ADD{$\lambda$}\DEL{damping coefficients} \ADD{for a large range of frequencies but the measurements are accurate only }\DEL{in a range of frequencies located }around the characteristic frequency of the pendulum, $\omega_c$.  To \DEL{get $C^{\mu}$}\ADD{increase the precision of the measurements of $\mu$} and \DEL{$C^{\lambda}$}\ADD{$\lambda$} as a function of the frequency, the different measurements performed at different positions of the counter-weight\ADD{s} are multiplied by a weight centered around the characteristic frequency $\omega_c$. Thus, the different experiments are combined using the most reliable region of each measurement. 

\ADD{Always two series of experiments are performed, one in a homogeneous fluid first and then one in a stratified fluid. In the case of a homogeneous fluid, $\mu$ is supposed to be independent of the frequency and $\lambda=\lambda_{\textrm{h}}$ is expected to be proportional to $\sqrt{\omega}$~\citep{Stokes1851,LandauLifshitz1959}. The inertial coefficient $C^{\mu}$ is obtained by averaging $\mu$ in the sampled frequency range and using equation~(\ref{eq:flat_plate_mu}). The viscous damping is quantified by a linear fit of $\lambda_{\textrm{h}}$ as a function of $\sqrt{\omega}$. For the stratified case, the inertial coefficient $C^{\mu}$ as a function of frequency is obtained using equation~(\ref{eq:flat_plate_mu}). The wave damping rate $\lambda_{\textrm{w}}$ is obtained by subtracting $\lambda_{\textrm{h}}$, previously measured in the homogeneous case, to $\lambda$. Then, the wave damping $C^{\lambda}$ and radiated wave power $P_{\textrm{w}}$ coefficients are computed using equations~~(\ref{eq:flat_plate_lambda}) and~(\ref{eq:power_coefficient}).} \DEL{To reduce the noise-over-signal ratio, three {impulse response functions} were recorded for each position of the counter-weight. Each measurement appears to be very reproducible.} \DEL{Moreover}\ADD{Finally}, using the \textit{medfilt1.m} Matlab function, we applied a median filter to $C^{\mu}$ and $C^{\lambda}$ \DEL{spectra }in order to slightly smooth  the curves.  The median filter has been performed on intervals of frequency $\omega$ of $0.05$~rad/s.

Note that the impulse response function analysis requires that the system is causal. Indeed, once the waves are emitted, they must not come back to the pendulum. Nevertheless, as the tank is limited in space, some waves are reflected at the ends of the tank and come back to the pendulum. Thus, they act as a source of oscillations and perturb the signal recorded. The wave-absorbers, placed at each end of the tank to prevent the internal waves to return to the pendulum, are not fully efficient but limit the energy coming back to the pendulum. Note that wave reflections are more important when the frequency $\Omega$ is low, i.e. when the waves propagate almost horizontally. In the following sections, to help the reader to identify quickly the range of frequencies where wave reflections are non-negligible in the experiments, the points showing the \DEL{added mass}\ADD{inertial}\DEL{ and}\ADD{, wave} damping \ADD{and radiated wave power} coefficients are empty when reflections are important and filled when reflections are small or nonexistent. \ADD{The importance of the reflections is determined by using the different individual records $x(t)$, not shown in this paper.}

\subsection{Set-up and data processing validation\label{validation}}

\ADD{Similar set-up and data-processing method have been used in the previous works by~\cite{Ermanyuk2000,Ermanyuk2002} and~\cite{ErmanyukGavrilov2002a,ErmanyukGavrilov2002b,ErmanyukGavrilov2003}. However, as the set-up and the method are similar but not strictly identical, we wanted to test them using previously published results. In order to do that, the first experiments have been performed using a cylinder with a square cross-section oscillating in a stratified fluid of large depth. Such experiments have already been reported by~\cite{ErmanyukGavrilov2002b}. In Appendix~\ref{square}, we compare our results and the ones obtained in~\cite{ErmanyukGavrilov2002b}, showing a good agreement between the two sets of data. We also correct a small error in calculations present in~\cite{ErmanyukGavrilov2002b}. \DEL{This demonstrates the ability of our current set-up to measure inertial and wave damping coefficients of an object oscillating in a stratified fluid of large depth.} In the reminder of the manuscript, the set-up is used to address the effects of finite depth on tidal conversion.}

\section{Circular cylinder\label{disk}}

In this section, we discuss the finite depth effects on \DEL{added mass}\ADD{inertial} and \ADD{wave} damping coefficients for a cylinder with a circular cross section oscillating horizontally in a stratified fluid. The \DEL{experimental parameters}\ADD{different characteristics of this shape} are given in Table~\ref{tabular:shape_parameters} \ADD{while the experimental parameters for the series of experiments can be found in Table~\ref{tabular:cylinder_parameters}}. As mentioned in the Introduction, the results from~\cite{ErmanyukGavrilov2002a} at $H/b$ from $4.32$ to $1.65$ show \ADD{for $\Omega>0.2$} a qualitative agreement with the theoretical behaviour predicted for subcritical obstacles~\citep{LlewellynSmithYoung2002} rather than for the supercritical ones~\citep{LlewellynSmithYoung2003}. The complementary set of data obtained at $H/b$ from $3.2$ to $1.2$ is described in this section. This data set extends the results to lower values of $H/b$. In addition, in this new data set the diameter of the cylinder is $b=5$~cm instead of $b=3.7$~cm used in~\cite{ErmanyukGavrilov2002a}\ADD{. This decreases the}\DEL{what implies a decreased} role of the \ADD{viscous} boundary layers on the results. Also, the free surface in the new set of experiments is replaced by a rigid lid, what ensures the identical conditions at the upper and lower boundaries of the fluid domain and also a sufficient length of the fluid domain to minimize the role of the internal wave reflections at the end walls. In~\cite{ErmanyukGavrilov2002a}, the presence of free surface forced the authors to limit the length of the test tank in order to avoid the excitation of the surface seiche within the frequency domain of interest. \ADD{This caused a limitation at small frequencies in the measurements. By adding a rigid lid and performing experiments with smaller $H/b$, we expect to reduce this limitation in frequency.} \ADD{Therefore, this study is focused on small frequencies where the cylinder is essentially in the supercritical regime and where enhancement of tidal conversion is expected.}
 
\begin{table}
\begin{center}
\begin{tabular}{c||c|c|c|c|c}
 Series & $H$~[cm]  & $H/b$ & $C^{\mu}$  &  $\lambda_{\textrm{h}}/\sqrt{\omega}$~[kg/s$^{1/2}$] & Symbols\\
 \hline
 $1$ & $16$ & $3.2$  & $1.39\pm0.05$ & $0.07\pm0.01$ & Green diamonds\\
 $2$ & $10$ & $2$  &  $1.75\pm 0.08$ & $0.10\pm0.01$ & Blue pentagrams\\
 $3$ & $7.5$ & $1.5$  & $2.46\pm0.15$ & $0.17\pm0.01$ & Red squares\\
$4$& $6$ & $1.2$ & $5.12\pm0.20$ & $0.67\pm0.02$ & Black circles\\
\end{tabular}

\caption{Parameters \DEL{and measured quantities }for the four series of experiments at finite depth and using the cylinder with a circular cross section. \ADD{The inertial coefficient $C^{\mu}$ and the viscous damping rate $\lambda_{\textrm{h}}$ are measured in the homogeneous case.} The symbols mentioned in the last column are used in figures~\ref{fig:mu_homogene_pendule_cylindrique},~\ref{fig:mu_pendule_cylindrique} and~\ref{fig:lambda_pendule_cylindrique}.}
\label{tabular:cylinder_parameters}
\end{center}
\end{table}

\subsection{Homogeneous fluid}

The solution for the {added mass} coefficient of an elliptic cylinder submerged at mid-depth of a horizontal \DEL{stripe}\ADD{channel} of homogeneous fluid has been obtained by~\cite{Clarke2001}
\begin{equation}
k_{\ast}\left(q,\xi\right)=2\frac{\ln\left[\sec \left(\frac{\pi q}{2}\frac{1}{1-\xi}\right)\right]}{\left(\frac{\pi q}{2}\frac{1}{1-\xi}\right)^2},\label{eq:clarke2001}
\end{equation}
where the parameter $\xi=\xi(p,q)$ is to be determined from the following equation
\begin{equation}
\frac{2\xi }{\pi q}\ln\left[\sec\left(\frac{\pi q}{2}\frac{1}{1-\xi}\right)+\tan\left(\frac{\pi q}{2}\frac{1}{1-\xi}\right)\right]=\frac{1}{p}.\label{eq:clarke2001_xi}
\end{equation}
In our case, the cross section of the cylinder is 
 circular so that $p=1$. Using equations~(\ref{eq:clarke2001}) and~(\ref{eq:clarke2001_xi}), it is thus possible to compute the \DEL{added mass}\ADD{inertial} coefficients in a homogeneous fluid as a function of  $H/b$. Table~\ref{tabular:c_mu_clarke} gives the measured and predicted \DEL{added mass}\ADD{inertial} coefficients \ADD{$C^{\mu}$} for our four series of experiments and the five ones available in~\cite{Ermanyuk2000} and in~\cite{ErmanyukGavrilov2002a}. First, one can see that all the data sets are in a good agreement. Second, the prediction of~\cite{Clarke2001} works well for high values of $H/b$ and remains reasonably accurate when $H/b$ decreases. Note that the numerical calculations of~\cite{Sturova2001} lead to \DEL{added mass }values close to the ones predicted by equations~(\ref{eq:clarke2001})  and~(\ref{eq:clarke2001_xi}). \ADD{Note also that the affine similitude theory cannot be used for predicting the added mass coefficients in the stratified case for $\Omega<1$ because the prediction of~\cite{Clarke2001} is not an analytical function. For $\Omega>1$, the prediction for the inertial coefficient are shown in figure~\ref{fig:mu_pendule_cylindrique} using dashed-dotted lines, with different colors corresponding to different fluid depths (see Table~\ref{tabular:cylinder_parameters}).}
 
\begin{table}
\begin{center}
\begin{tabular}{c|l|c|c}
 $H/b$ & ~~~~~~~~~~~Reported in & $C^{\mu}$ measured & $C^{\mu}$ predicted~\citep{Clarke2001}\\
 \hline
$7.57$& \cite{Ermanyuk2000} &$1.05\pm0.05$  & $1.029$\\
$4.32$ &\cite{ErmanyukGavrilov2002a} &$1.12\pm0.05$  & $1.093$\\
$3.24$ &\cite{ErmanyukGavrilov2002a} &$1.24\pm0.05$  & $1.173$\\
$3.2$ & present paper &$1.39\pm0.05$  & $1.178$\\
$2.19$ &\cite{ErmanyukGavrilov2002a} &$1.54\pm0.05$  & $1.438$\\
$2$ & present paper &$1.75\pm0.08$  & $1.558$\\
$1.65$ &\cite{ErmanyukGavrilov2002a} &$2.25\pm0.09$  & $1.999$\\ 
$1.5$ & present paper &$2.46\pm0.15$  & $2.418$\\
$1.2$ & present paper &$5.12\pm0.20$  & $5.825$\\
\end{tabular}
\caption{Measured (third column) and predicted (fourth column) \DEL{added mass}\ADD{inertial} coefficients $C^{\mu}$ in a homogeneous fluid for the different $H/b$ ratios shown in the first column. }
\label{tabular:c_mu_clarke}
\end{center}
\end{table}

\begin{figure}
\begin{center}
\includegraphics[width=0.5\linewidth,clip=]{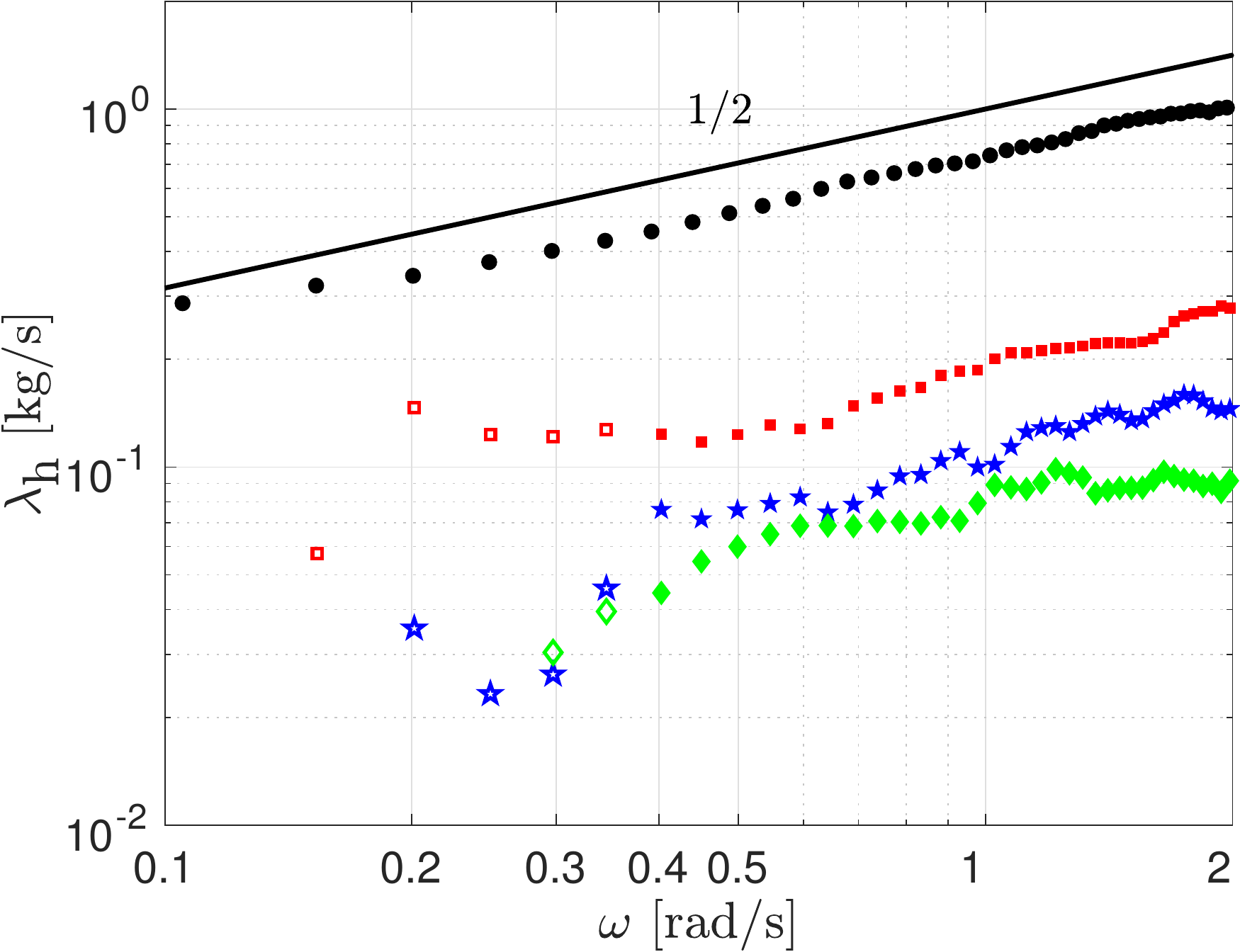}
    \caption{\ADD{Homogeneous case:} viscous damping \DEL{coefficient}\ADD{rate} $\lambda_{\textrm{h}}$ as a function of frequency $\omega$ for the four series of experiments performed with our pendulum set-up in log-log scale. $H/b$ is equal to $1.2$ (black circles), $1.5$ (red squares), $2$ (blue pentagrams) and $3.2$ (green diamonds).     The solid black line shows the power law $\lambda_{\textrm{h}}\propto \sqrt{\omega}$.\label{fig:mu_homogene_pendule_cylindrique}}
   \end{center}
\end{figure}

\DEL{It is necessary to calibrate the pendulum in a homogeneous fluid for the four different depths, as the viscous dissipation $\lambda_{\textrm{h}}$ changes with $H$. For the damping coefficient $\lambda_{\textrm{h}}$ in a homogeneous fluid, one expect the classic Stokes' dependence $\lambda_{\textrm{h}} \propto \sqrt{\omega}$.} Figure~\ref{fig:mu_homogene_pendule_cylindrique} shows the viscous damping \DEL{coefficients}\ADD{rates $\lambda_{\textrm{h}}$} for the four series of experiments performed with the present pendulum set-up as a function of the frequency $\omega$, in log-log scales. One can see that for frequencies higher than $\omega=0.5$~rad/s, the \ADD{viscous} damping \ADD{rate}\DEL{coefficient} $\lambda_{\textrm{h}}$ agrees reasonably well with the square root of the frequency $\omega$\ADD{~\citep{Stokes1851,LandauLifshitz1959}}. Below $\omega=0.5$~rad/s, the signal is noisy except for the smallest value of $H/b$ (black filled circles). The different values of \ADD{the coefficients} $\lambda_{\textrm{h}}/\sqrt{\omega}$ are reported in Table~\ref{tabular:cylinder_parameters}.

\subsection{Stratified fluid}

\begin{figure}
\begin{center}
\includegraphics[width=0.55\linewidth,clip=]{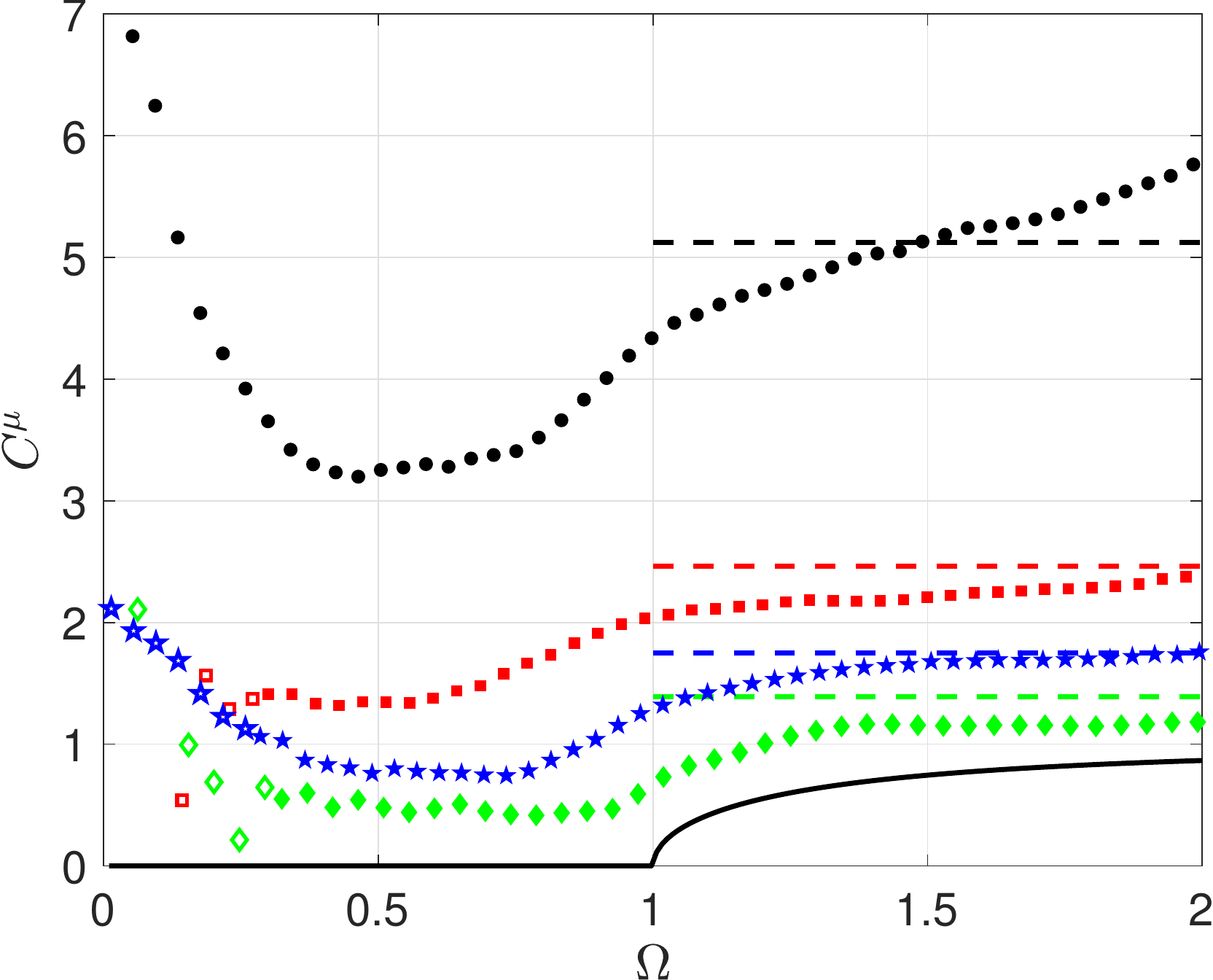}
    \caption{\ADD{Stratified case:} \DEL{Added mass}\ADD{inertial} coefficient $C^{\mu}$ as a function of the dimensionless frequency $\Omega$ for the four different series of experiments at different depths. $H/b$ is equal to $1.2$ (black circles), $1.5$ (red squares), $2$ (blue pentagrams) and $3.2$ (green diamonds). The frequency range where wave reflections are important corresponds to the range where the symbols are empty. The horizontal dashed lines correspond to the \DEL{added mass}\ADD{inertial coefficient} measured in a homogeneous fluid for the different depths. The solid black line corresponds to the theoretical prediction at infinite depth~\citep{Hurley1997} \ADD{and the dash-dotted lines represent the prediction made using the affine similitude theory, for $\Omega>1$}. The color code for lines is identical to the one for the symbols.\label{fig:mu_pendule_cylindrique}}
   \end{center}
\end{figure}

Figure~\ref{fig:mu_pendule_cylindrique} shows the \DEL{added mass}\ADD{inertial} coefficients \ADD{$C^{\mu}$} for the four series of experiments. The \DEL{added mass} prediction \ADD{made by}~\cite{Hurley1997} for a fluid of infinite depth is represented by the solid black line \ADD{and the predictions from the affine similitude theory for finite depth are shown using the dashed-dotted lines}. Note that for $\Omega<1$, the predicted \DEL{added mass}\ADD{inertial} coefficient is identically zero \ADD{for a fluid of infinite depth}.
In figure~\ref{fig:mu_pendule_cylindrique}, one can see that the different \DEL{added mass}\ADD{inertial} coefficients reach some asymptotic values for $\Omega>1$, except for the smallest $H/b$ ratio (black circles). These asymptotic values are close to the \DEL{added mass }values measured in a homogeneous fluid, indicated by dashed lines. This is consistent with the prediction that, at high frequencies, the \DEL{added mass}\ADD{inertial} coefficients are not different from the ones of the homogeneous case. \ADD{The measurements are also in a relatively good agreement with the prediction of the affine similitude theory, except for $\Omega$ close to $1$ where $\alpha$ tends to $0$. Even at $\Omega>1$  the solution does not work well for vertically squeezed geometries, and therefore cannot be effectively used when $b/a=p$ is $O(1)$ and $H/b \rightarrow 1$.} One can note that as $H/b$ decreases, the \DEL{added mass}\ADD{inertial} coefficient increases. Thus, the deeper, the smaller the \DEL{added mass}\ADD{inertial coefficient}, at any frequency. This is consistent with the measurements performed by~\cite{ErmanyukGavrilov2002a}. At low frequency,  the results are perturbed (see empty symbols) by wave reflections. Note that the range of frequency where the signal is noisy is reduced when one decreases the depth. Indeed, a small depth imposes to almost all waves to be reflected a given number of times at the top and bottom of the tank before reaching the ends of the tank. These multiple reflections induce a significant decay of waves.

\begin{figure}
\begin{center}
\includegraphics[width=\linewidth,clip=]{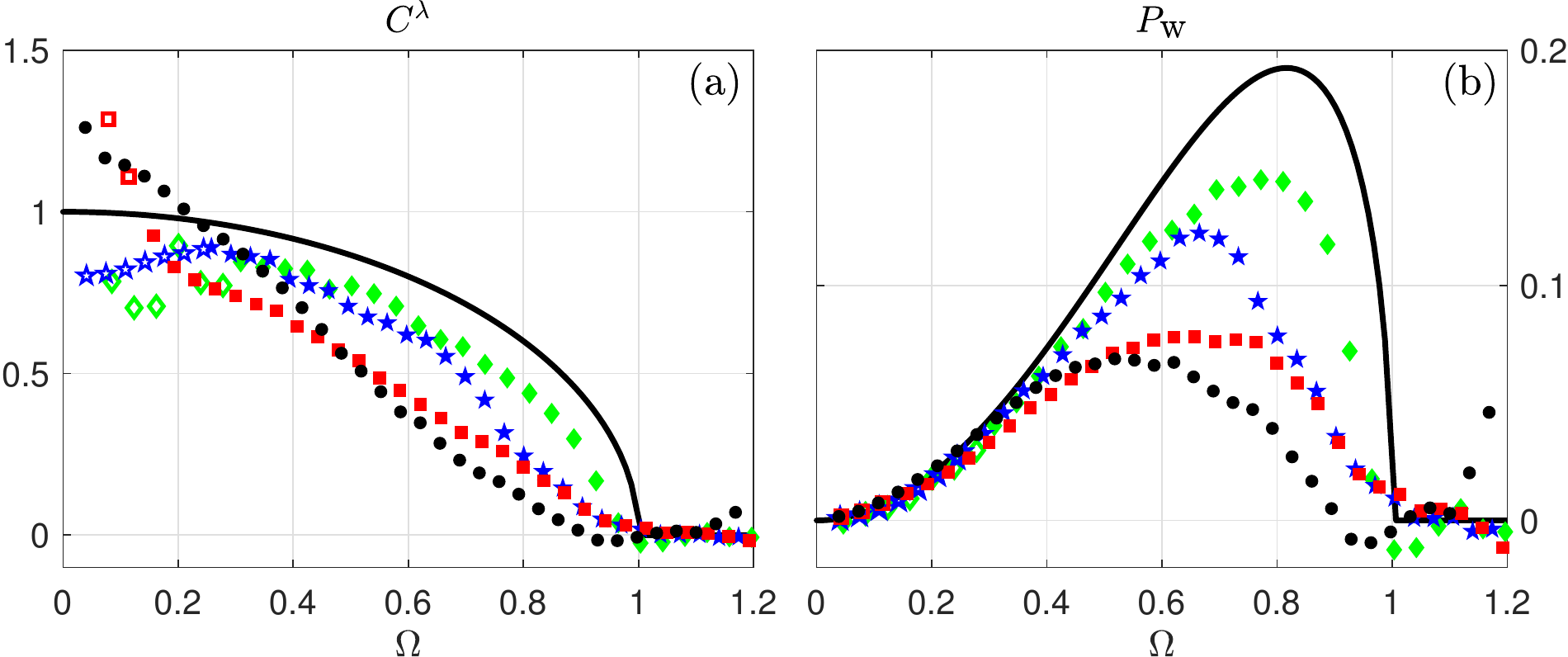}
    \caption{\ADD{Stratified case:} \ADD{wave} damping coefficient $C^{\lambda}$ (a) and radiated wave power $P_{\textrm{w}}$ (b) as a function of the dimensionless frequency $\Omega$ for the four different series of experiments at different depths. As in figure~\ref{fig:mu_pendule_cylindrique}, $H/b$ is equal to $1.2$ (black circles), $1.5$ (red squares), $2$ (blue pentagrams) and $3.2$ (green diamonds). The frequency range where wave reflections are important corresponds to the range where the symbols are empty. The solid black line corresponds to the theoretical prediction at infinite depth~\citep{Hurley1997}.\label{fig:lambda_pendule_cylindrique}}
   \end{center}
\end{figure}

Figure~\ref{fig:lambda_pendule_cylindrique} shows the \ADD{wave} damping \ADD{$C^{\lambda}$} and radiated wave power \ADD{$P_{\textrm{w}}$} coefficients for the four series of experiments. The symbols used are the same as in figure~\ref{fig:mu_pendule_cylindrique} and are described in Table~\ref{tabular:cylinder_parameters}. The prediction for a fluid of infinite depth~\citep{Hurley1997} is plotted as a solid black line, on both panels. For $\Omega>1$, one can see that no wave is emitted and $C^{\lambda}$ vanishes for all the four cases. In the studied range of $H/b$, the efficiency of wave radiation in the frequency range $0.5<\Omega<1$ drops systematically as the depth of fluid decreases. This behaviour is in qualitative agreement with~\cite{LlewellynSmithYoung2002} for a subcritical obstacle in a fluid of limited depth, and also with~\cite{GorodtsovTeodorovich1986} and~\cite{ErmanyukGavrilov2002a}. {However, at lower frequency, the opposite trend can be also observed~\citep{LlewellynSmithYoung2003}: there is an enhancement of wave radiation at $H/b=1.2$ and $1.5$ at $\Omega\rightarrow 0$ as expected for a supercritical obstacle, although it has a small magnitude in the case of a circular cylinder. 

\ADD{For a vertical plate, a fully supercritical case, o}\DEL{O}ne can expect that the force coefficients of a vertical plate \ADD{are enhanced} \DEL{follow similar trends} at $H/b\rightarrow 1$ \DEL{and $\Omega\rightarrow 0$}\ADD{for all frequencies $\Omega$}. This issue is addressed in the next section. 

\section{Vertical plate\label{vertical_plate}}

In this section, we present  \DEL{added mass}\ADD{inertial} and \ADD{wave} damping coefficients of a vertical plate oscillating horizontally in a stratified fluid of limited depth. To our knowledge, no \ADD{finite-depth} experiment with such a shape has been performed before. \ADD{However, it is worthwhile to note that~\cite{Peacocketal2008} have reported experiments with a knife edge in the infinite-depth limit, without measuring the inertial and radiated wave power coefficients.} The characteristics of the plate are given in Table~\ref{tabular:shape_parameters}. Its vertical size is equal to $b=10$~cm and it has a thickness of $a=0.3$~cm. Thus, $b\gg a$  and $p$ is large (order 30). The cross section of this vertical flat plate can also be seen as an ellipse with a very small horizontal scale $a$.

Two series of experiments have been performed, with two different depths, corresponding to $H/b=2.2$ and $H/b=1.3$. These $H/b$ ratios have been chosen to be small, in order to observe a similar enhancement of the \ADD{wave} damping coefficient at small depth and frequency as the one revealed in section~\ref{disk} for the cylinder with a circular cross section. Table~\ref{tabular:vertical_plate_parameters} gives the different parameters for the two series of experiments.

\subsection{Theoretical predictions}

The added mass of a flat plate oscillating in a homogeneous fluid of infinite depth ($H \rightarrow \infty$ and $q=b/H \rightarrow 0$) is $m_A=\rho_0 \pi b^2/4$ (see for example~\cite{Brennen1982}). Therefore, the function given in equation~(\ref{functions}) is $f_{\ast}(\infty, 0)\equiv 1$. 
Thus, using the affine similitude theory, the \DEL{added mass}\ADD{inertial} and \ADD{wave} damping coefficients in a linearly stratified fluid of infinite depth coincide with the solution found by~\cite{Hurley1997}, also valid for the cylinder having a circular cross section (see section~\ref{disk}). Consequently, the non-dimensional wave power radiated by the flat plate in a uniformly stratified fluid of infinite extent amounts to
\begin{equation}
P_{W}(\Omega,q=0)=\frac{1}{2}\Omega^2 (1-\Omega^2)^{1/2}.
\label {power1}
\end{equation}
Note that the same result is true at finite $p=b/a$ since the added mass of an elliptic cylinder in the fluid of infinite extent depends only on its vertical size and does not depend on its elongation. This is in full qualitative agreement with~\cite{Bell1975}.

Following~\cite{LockwoodTaylor1930}, the {added mass} coefficient of a vertical flat plate of height $b$ oscillating in a homogeneous fluid of finite depth $H$ is
\begin{equation}
k_{\ast}=f_{\ast} \left(\infty, q\right)=2 \times \left(\frac{2}{\pi q}\right)^2 \ln\left[\sec\left(\frac{\pi q}{2}\right)\right].
\label {addedmass}
\end{equation}
Thus, when $\Omega<1$, the wave power radiated by the flat plate in a uniformly stratified fluid of limited depth is
\begin{equation}
     P_{W}(\Omega,q)=\frac{1}{2}\Omega^2 (1-\Omega^2)^{1/2} \mathcal{E} \left(q\right),
    \label {power2}
   \end{equation}
where 
 we introduce the enhancement factor
\begin{equation}
\mathcal{E}\left(q\right)=\frac{P_{W}(\Omega,q)}{P_{W}(\Omega,q=0)}=\frac{f_{\ast} (\infty, q)}{f_{\ast} (\infty, 0)}=f_{\ast} \left(\infty, q\right).\label{eq:enhancement_factor}
\end{equation}
Thus, the \DEL{added mass}\ADD{inertial} and \ADD{wave} damping coefficients of a vertical plate oscillating horizontally in a fluid on finite depth are \ADD{expected to be} equal to the ones for a fluid of infinite depth multiplied by the enhancement factor~$\mathcal{E}$. The above expression for the enhancement factor is equivalent to the result given by a more complicated formula in~\cite{LlewellynSmithYoung2003} \ADD{(see equations (4.7) and (4.9) of this reference)}, with rotation neglected
and taking into account a factor 2 
since they consider only the upper half of geometry. \ADD{Explicitly, it means that
\begin{equation}
\frac{2}{\pi}\left[1-\tan^{2}\left(\frac{\pi q}{2}\right)\right]^{1/2}\int_{0}^{\tan(\frac{\pi q}{2})}\frac{\xi\arctan(\xi) d\xi}{\left[\tan^2\left(\frac{\pi q}{2}\right)-\xi^{2}\right]^{1/2}(1-\xi^{2})}=\ln\left[\sec\left(\frac{\pi q}{2}\right)\right]
\end{equation}}
It is worth to note that 
the wave power given in~\cite{LlewellynSmithYoung2003} is in the low-frequency limit $\Omega\rightarrow 0$, where the dimensionless \ADD{wave} damping coefficient is given by
\begin{equation}
C^{\lambda}(\Omega\rightarrow 0)=\mathcal{E}\left(q\right).
\end{equation}
When $H \rightarrow \infty$, we have $q \rightarrow 0$ and $\mathcal{E} \rightarrow 1$. However, for smaller depths (or larger $q$), $\mathcal{E}$ becomes greater than $1$, which highlights a more important \ADD{wave} damping coefficient at finite depth than at infinite depth, in the low frequency limit \DEL{useful for tidal conversion}\ADD{relevant for the supercritical regime}.
Note also that the enhancement factor is the same both for the \DEL{added mass}\ADD{inertial coefficient $C^{\mu}$} at $\Omega \rightarrow \infty$ and for the \ADD{wave} damping coefficient \ADD{$C^{\lambda}$} at $\Omega \rightarrow 0$.
Following this theory,  the \DEL{solid curves for the added mass}\ADD{inertial coefficient predictions} are presented in figure~\ref{fig:flatplate_addedmass}\ADD{, using solid lines while in figure~\ref{fig:flatplate_damping}(a), we show the theoretical prediction for the \ADD{wave} damping coefficient.}

It it also possible to use the prediction of~\cite{Clarke2001} by considering that the vertical plate has a shape close to a very elongated ellipse. As the thickness of the plate is very small, we have to take into account the viscous boundary layers, whose thickness $\delta$ can be of the same order of magnitude as the one of the plate. It is given by
\begin{equation}
\delta(\omega)=\sqrt{\frac{2\nu}{\omega}},
\end{equation}
where $\omega$ is the frequency of the oscillations and $\nu$ the kinematic viscosity of the fluid. For frequencies of the order of $1$~rad/s, one gets $\delta\approx 0.2$~cm, which is close to the thickness of the plate. The theoretical curves for the \DEL{added mass}\ADD{inertial coefficient} obtained using this method are represented in figure~\ref{fig:flatplate_addedmass} for $\Omega>1$ by dashed dotted lines. 
One can clearly see that the anticipated effect of the finite thickness of the vertical flat plate is low even if we add the "virtual" boundary layer thickness 
on two sides of the plate $2\delta$ to the actual thickness of the plate $a$. \DEL{Also, in figure~\ref{fig:flatplate_damping}(a), we show the theoretical prediction for the \ADD{wave} damping coefficient for the flat plate in the fluid of finite depth. N}\ADD{As in section~\ref{disk}, n}ote that it does not seem possible to continue~(\ref{eq:clarke2001}) and~(\ref{eq:clarke2001_xi}) to $\Omega<1$ since the solution is not given by an explicit function. \DEL{Also even at $\Omega>1$  the solution does not work well for vertically squeezed geometries, and therefore cannot be effectively used when $b/a=p$ is $O(1)$ and $H/b \rightarrow 1$.}

\subsection{Experimental results}

Table~\ref{tabular:vertical_plate_parameters} summarizes the results obtained in a homogeneous fluid for the vertical plate at two different depths. Similar to the case of a circular cylinder, the \DEL{added mass}\ADD{inertial coefficient $C^{\mu}$} and viscous damping \ADD{rate $\lambda_{\textrm{h}}$} increase as the depth decreases. The measured values of the \DEL{added mass}\ADD{inertial coefficient} are typically $10$ to $20\%$ larger than the predicted ones. 

\begin{table}
\begin{center}
\begin{tabular}{c||c|c|c|c|c|c}
 Series & $H$~[cm]  & $H/b$ & $C^{\mu}$  & $C^{\mu}$ predicted &  $\lambda_{\textrm{h}}/\sqrt{\omega}$~[kg/s$^{1/2}$] & Symbols\\
 \hline
 $1$ & $22$ & $2.2$  & $1.24\pm 0.1$ & $1.10$ & $0.24\pm0.03$ & Blue pentagrams\\
 $2$ & $13$ & $1.3$  & $1.79\pm0.07$ & $1.42$ & $0.39\pm0.03$ & Red squares\\
\end{tabular}
\caption{Parameters\DEL{, measured quantities and predicted added masses} for the two series of experiments at finite depth and using the vertical flat plate as the cylinder. \ADD{The inertial coefficient $C^{\mu}$ and the viscous damping rate $\lambda_{\textrm{h}}$ are measured in the homogeneous case.} The prediction for the \DEL{added mass}\ADD{inertial coefficient in the homogeneous case} is obtained from equation~(\ref{eq:enhancement_factor}). The symbols mentioned in the last column are used in figures~\ref{fig:flatplate_addedmass} and~\ref{fig:flatplate_damping}.}
\label{tabular:vertical_plate_parameters}
\end{center}
\end{table}

Figure~\ref{fig:flatplate_addedmass} shows the \DEL{added mass}\ADD{inertial} coefficient \ADD{$C^{\mu}$} as a function of  $\Omega$ for the vertical flat plate oscillating horizontally in a stratified fluid of limited depth. The symbols used here are described in Table~\ref{tabular:vertical_plate_parameters}. For $H/b=2.2$ (blue pentagrams) and $\Omega>1$, the experimental and the two theoretical curves are close  to each other, particularly for large frequencies. For $H/b=1.3$, experimental points fall significantly above the theoretical curves. However, for both fluid depths, the \DEL{added mass}\ADD{inertial coefficient} does not vanish for $\Omega<1$, contrary to the prediction.  The \DEL{added mass}\ADD{inertial} coefficient as a function of the frequency shows a very similar behaviour to the one for the circular cylinder (see figure~\ref{fig:mu_pendule_cylindrique}). Note that, as in section\DEL{s~\ref{square} and}~\ref{disk}, data are less reliable at small frequency due to the wave reflections. This is highlighted by empty symbols. 

\begin{figure}
\begin{center}
\includegraphics[width=0.5\linewidth,clip=]{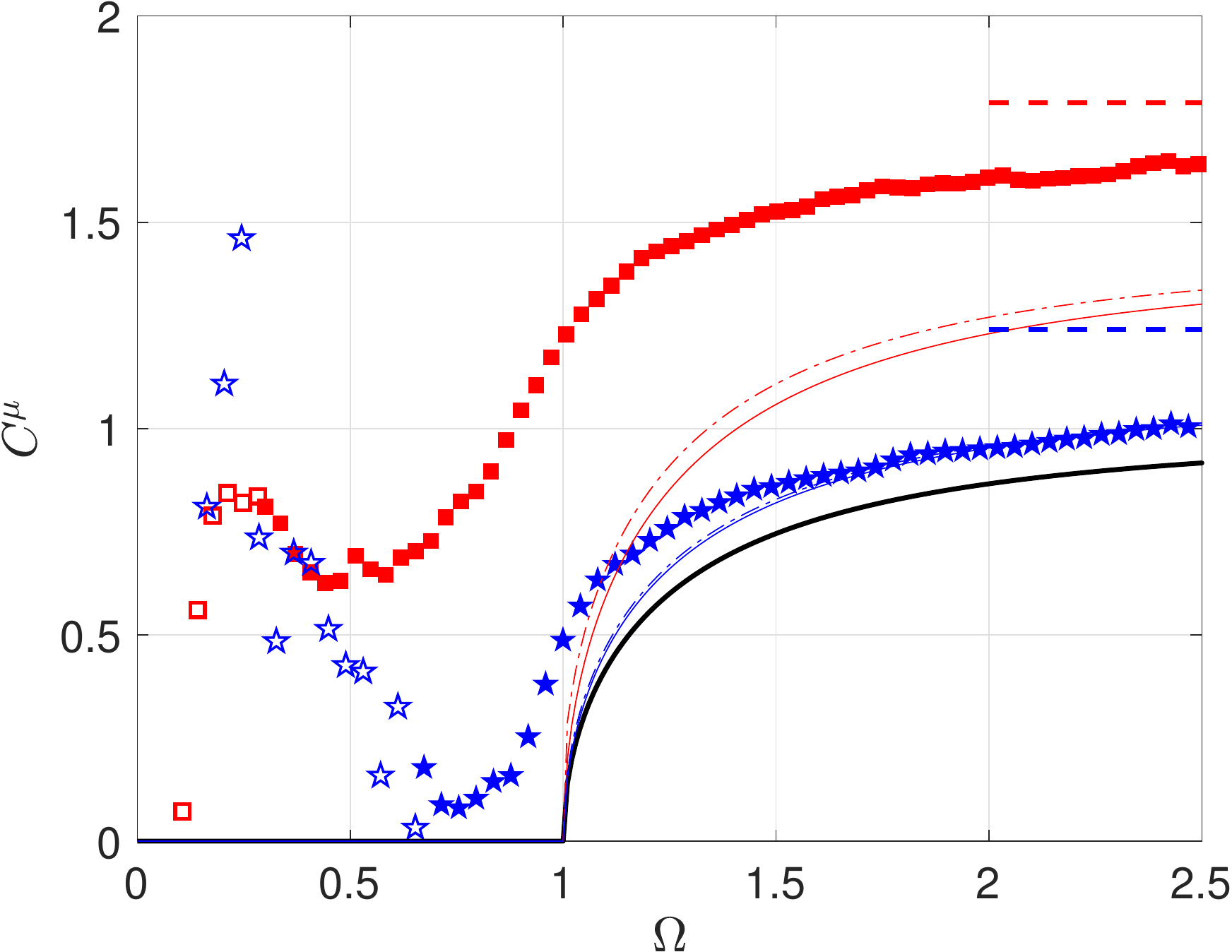}
    \caption{\ADD{Stratified case:} \DEL{Added mass}\ADD{inertial} coefficient $C^{\mu}$ as a function of the dimensionless frequency $\Omega$ for the vertical flat plate at two different fluid depths. $H/b$ is equal to $2.2$ (blue stars) and $1.3$ (red squares). The frequency range where wave reflections are important corresponds to the range where the symbols are empty. The horizontal dashed lines correspond to the \DEL{added mass}\ADD{inertial coefficient} measured in a homogeneous fluid for the two depths while the solid lines correspond to the theoretical prediction for a vertical plate and the dashed dotted lines to the theoretical prediction for a very thin ellipse. The colors are the same than the ones for the symbols. The black line represents the solution of~\cite{Hurley1997}, in a fluid of infinite depth. \label{fig:flatplate_addedmass} }
   \end{center}
\end{figure}

\begin{figure}
\begin{center}
\includegraphics[width=0.95\linewidth,clip=]{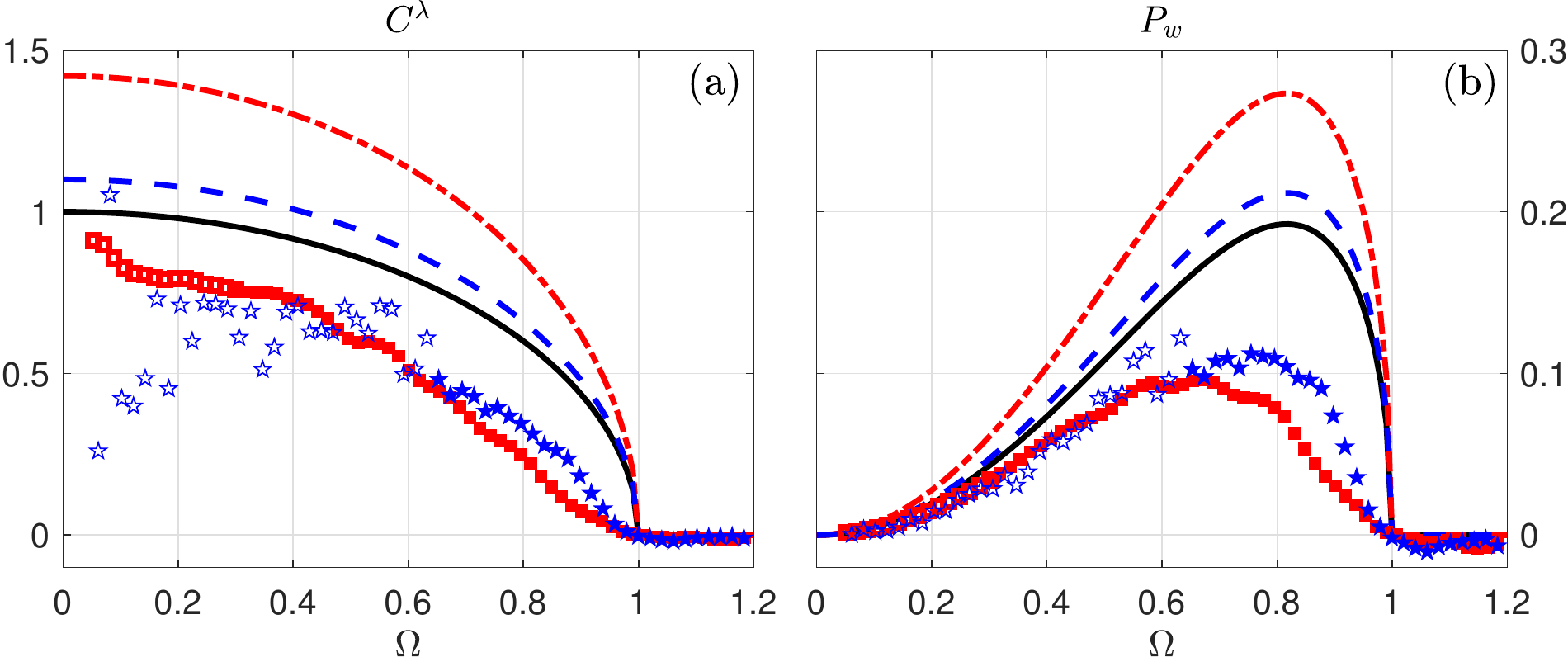}
    \caption{\ADD{Stratified case:} \ADD{wave} damping coefficient $C^{\lambda}$ (a) and radiated power $P_{\textrm{w}}$ (b) as a function of the dimensionless frequency $\Omega$ for the vertical flat plate at two different depths. As in figure~\ref{fig:flatplate_addedmass}, $H/b$ is equal to $2.2$ (blue stars) and $1.3$ (red squares). The frequency range where wave reflections are important corresponds to the range where the symbols are empty. The solid black line corresponds to the theoretical prediction at infinite depth made by~\cite{Hurley1997} while the two coloured lines are obtained by multiplying this prediction by the enhancement factor corresponding to the two different series of experiments, according to equation~(\ref{addedmass}).  \label{fig:flatplate_damping}}
   \end{center}
\end{figure}

Figure~\ref{fig:flatplate_damping} shows the \ADD{wave} damping \ADD{$C^{\lambda}$} and radiated wave power \ADD{$P_{\textrm{w}}$} coefficients as a function of $\Omega$ for the vertical flat plate. The symbols are the same as the ones used in figure~\ref{fig:flatplate_addedmass} and are described in Table~\ref{tabular:vertical_plate_parameters}. The enhancement factor given in equation~(\ref{eq:enhancement_factor}) predicts that the \ADD{wave} damping coefficient and the radiated wave power at finite depth (colour curves) are larger than the ones at infinite depth (solid black line), for all $\Omega<1$. This is obviously not the case here because both curves are below the prediction made by~\cite{Hurley1997} for a fluid of infinite depth. Thus the cases of a flat plate and a circular cylinder show a similar behaviour (compare figures~\ref{fig:lambda_pendule_cylindrique} and~\ref{fig:flatplate_damping}) \ADD{even if the two bodies have different criticality}. 
One can expect a \ADD{wave} damping coefficient larger than~$1$ at $H/b\rightarrow 1$ and $\Omega\rightarrow 0$, as predicted by~\cite{LlewellynSmithYoung2003} and seen in section~\ref{disk}. This is not the case here, despite a small ratio $H/b=1.3$.
It is difficult experimentally to go beyond and decrease this ratio. \DEL{As in section~\ref{disk}, we are limited to $H/b\approx 1.2-1.3$.} In the studied range of parameters we observed no enhancement for the \ADD{wave} damping coefficient of a flat plate at $\Omega<1$, \ADD{despite the supercritical situation. T}\DEL{while t}he enhancement is \ADD{however} fully present for the \DEL{added mass}\ADD{inertial coefficient} at $\Omega>1$. 

As a plausible explanation, we \ADD{may} attribute the apparent absence of enhancement in \ADD{wave} damping coefficient in our experiments to the scale effect. Indeed the ideal-fluid theory (see figure~1 in~\cite{LlewellynSmithYoung2003}) considers the flow scheme with infinitely thin beams emanating from the tip of the vertical plate. The upward-emitted beams are reflected at the horizontal boundary and pass parallel and at small distance $2\delta \cos\theta$ from the downward-emitted beams, where $\delta$ is the height of the gap between the tip of the plate and the horizontal boundary. At laboratory scale the width of each beam in a viscous fluid is finite. This width can exceed the inter-beam distance $2\delta \cos\theta$ as $H/b\rightarrow 1$. Therefore the beams can overlap, \DEL{what completely changes the flow structure, and} leaving a possibility of destructive interactions.\DEL{between the beams.}\DEL{ The relevance of the ideal-fluid solution~\citep{LlewellynSmithYoung2003}} \ADD{A possible consequence for}\DEL{ to} large-scale objects \ADD{remains unclear}\DEL{ is also not clear} since the\DEL{ finite} width of beams at natural scale can be governed by turbulent viscosity \ADD{(see for example~\cite{Coleetal2009})}.

\section{Flattop hill: topography lacking of tidal conversion\label{flattop}}

In this last section, we present experiments performed with an object having a specific shape. This object has been designed thanks to Leo Maas in order to show \ADD{experimentally} that, at finite depth and for a given frequency, the tidal conversion vanishes. \ADD{Using an integral-equation method to compute the tidal conversion of a triangular ridge in a stratified fluid of finite depth, ~\cite{Petrelisetal2006} have previously reported an example of a topography lacking of tidal conversion. Then,} \DEL{Topographies lacking of tidal conversion have indeed been found theoretically by}~\cite{Maas2011} \ADD{has shown that a whole family of such topographies exists.} \DEL{and a}\ADD{More recently,} a series of experiments have been reported \DEL{recently }showing such lack of tidal conversion~\citep{Pacietal2015}. \ADD{However, the radiated wave power has never been measured experimentally.} \DEL{By measuring the radiated wave power, w}\ADD{Here, w}e show a significant decay of the wave power for a ``perfect tuning" of geometry, depth and frequency, and how it is affected by detuning.

\subsection{The shape of the object and theoretical prediction}

Figure~\ref{fig:flattop_shape}(a) shows the shape of the object in the $x-z$ plane. Its height $b$ is equal to $8$~cm and it has a horizontal length $a$ of $20$~cm. The object is invariant in the $y$-direction. As the shape is symmetric with respect to the $x$ and $z$ axes, it is composed of one specific curve shown in figure~\ref{fig:flattop_shape}(b). The curve has a flat part for $x<2.26$~cm, with the limit marked by a blue circle. Then there is a decay with an inflection point, marked using a blue dot. 
At this inflection point, the slope has an angle of $36.6^\circ$ which corresponds to frequency of $\Omega_s=0.596$ for internal waves. This slope is shown by a blue dashed line in figure~\ref{fig:flattop_shape}(b). The inflection point has, by definition, the steepest slope of the curve. Another blue circle shows the end of the curve, close to $x=10$~cm. The two blue circles and the blue dot are shown on each side of the shape in figure~\ref{fig:flattop_shape}(a).

\cite{Maas2011} predicted that such a shape can exhibit a lack of tidal conversion for given frequency and depth. Indeed, if the depth is fixed, there can exist a specific frequency $\Omega_{\ell}$ such that the two sides of the shape are connected by rays after a reflection at the surface or at the bottom. This is illustrated in figure~\ref{fig:flattop_shape}(a) by the red lines showing internal wave rays and the dashed blue lines representing the surface and the bottom of the tank. With this geometry and at this specific frequency, the internal wave beams emitted by the pendulum are annihilated by pairs, after reflection on the surface or at the bottom of the tank, leading to a lack of tidal conversion.  When the fluid depth $H$ varies, the frequency corresponding to the lack  of tidal conversion $\Omega_{\ell}$ also changes. Figure~\ref{fig:flattop_prediction} shows the height of the fluid $H$ as a function of the frequency $\Omega_{\ell}$. This frequency 
is much smaller than $1$ for small depth. When $H \rightarrow \infty$, the dimensionless frequency $\Omega_{\ell}$ shifts towards $1$, i.e. the limit of the wave emission frequency range. Note that this is valid only for the frequency higher than $\Omega_s$, i.e. when the shape is subcritical. Indeed, one can easily understand that for $\Omega<\Omega_s$, some internal wave rays should go through the object to connect the two sides of the shape. Thus, we should not observe a lack of wave emission for very small depths, for $H$ below $13$~cm. 

\begin{figure}
\begin{center}
\includegraphics[width=0.95\linewidth,clip=]{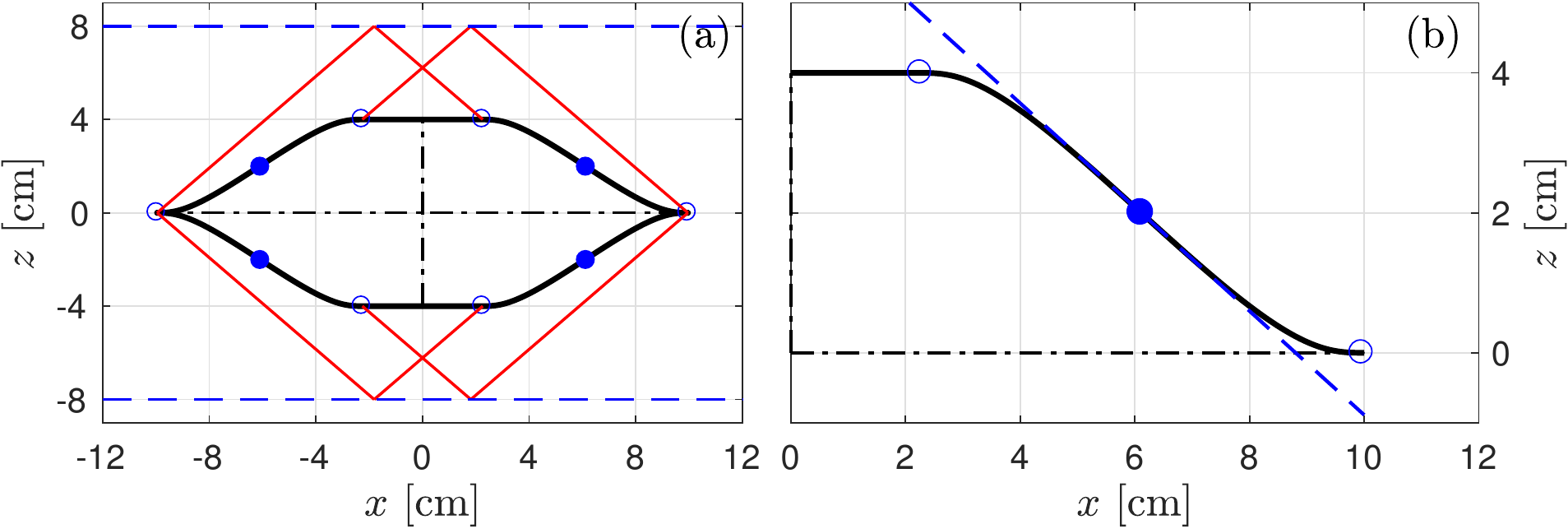}
    \caption{(a): Shape of the object in the $x-z$ plane (solid thick black line). The object is symmetric with respect to the $x$ and $z$ axes shown by the black dashed dotted lines inside. For this example, the cylinder is placed in the stratification of depth $H=16$~cm, limited by the two horizontal dashed blue lines. The red lines connect the two sides of the cylinder. (b): Zoom on the upper right quarter of the object. The inflection point is marked by the blue point. The slope of the shape at this point is shown by the dashed blue line. The two blue circles shows the portion of the shape which is not constant.\label{fig:flattop_shape}}
   \end{center}
\end{figure}

\begin{figure}
\begin{center}
\includegraphics[width=0.5\linewidth,clip=]{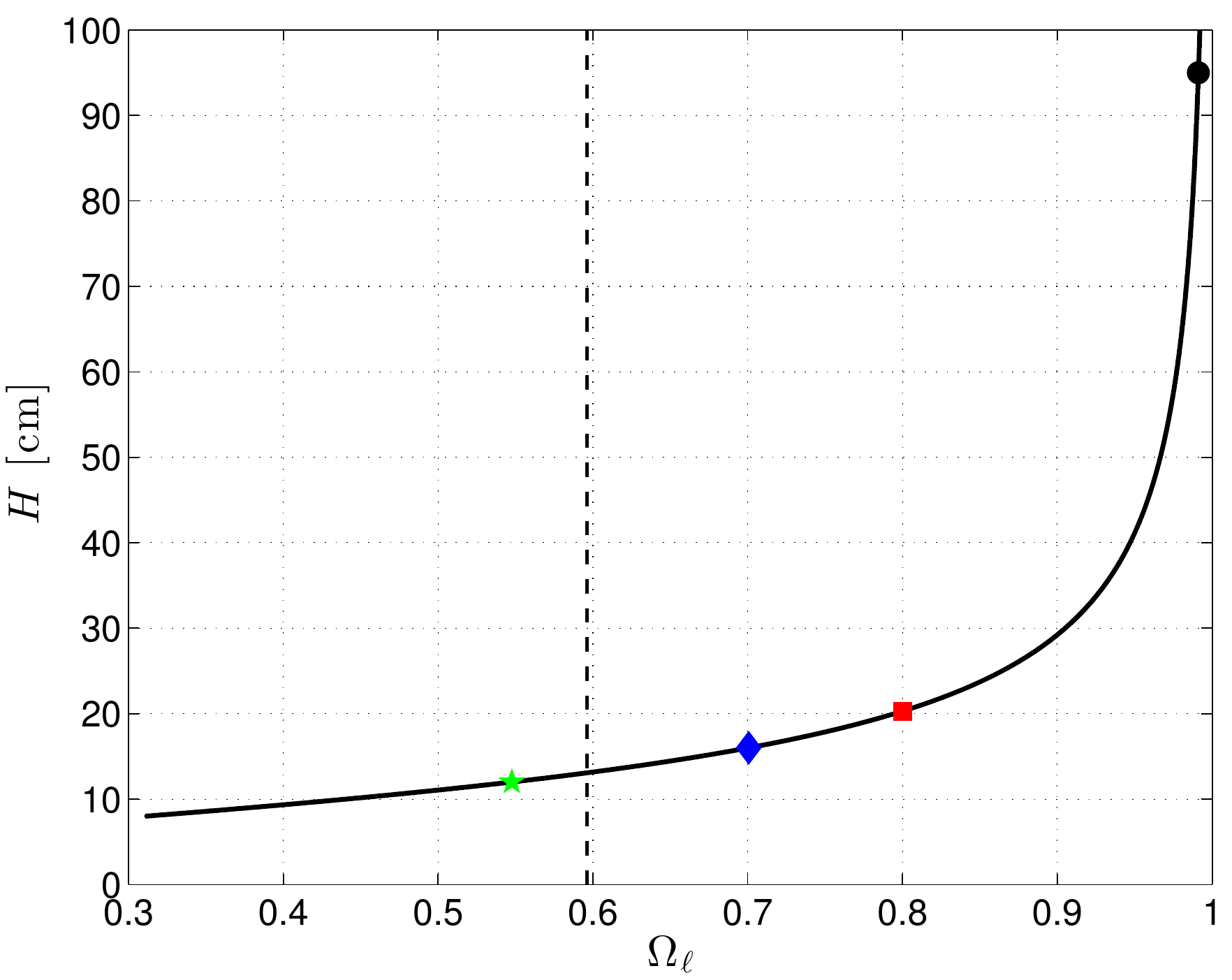}
    \caption{Correspondence between the height of the fluid $H$ and the frequency where the topography lacks of tidal conversion $\Omega_{\ell}$. The vertical dashed line shows the frequency $\Omega_s=0.596$. The four symbols represent the four different fluid depths reported in this paper.\label{fig:flattop_prediction}}
   \end{center}
\end{figure}

In the next section, we discuss four series of experiments performed at 
$H=95$, $20.3$, $16$ and $12$~cm. These experiments are shown in figure~\ref{fig:flattop_prediction} by the black circle, the red square, the blue diamond and the green pentagram, respectively. The prediction gives $\Omega_{\ell}$ equal to $0.99$, $0.8$, $0.7$ and $0.55$. 
Thus, we have explored the two different regions (subcritical and supercritical) in figure~\ref{fig:flattop_prediction}, separated by the dashed line at $\Omega=\Omega_s$. Before performing these four series of experiments, the cylinder with this flattop hill cross section has been calibrated in the air. The main characteristics are presented in Table~\ref{tabular:shape_parameters}. \ADD{Note that due to the very specific shape of the body, there is no theoretical prediction for the inertial and wave damping coefficients.}

\subsection{Experimental results}

\begin{table}
\begin{center}
\begin{tabular}{c||c|c|c|c|c}
 Series & $H$~[cm] & $H/b$ & $C^{\mu}$  &  $\lambda_{\textrm{h}}/\sqrt{\omega}$~[kg/s$^{1/2}$] & Symbols\\
 \hline
 $1$ & $95$ & $11.88$ & $0.78\pm0.10$ & $0.12\pm0.01$ & Black circles\\
 $2$ & $20.3$ & $2.54$ & $1.79\pm0.10$ & $0.20\pm0.01$ & Red squares\\
 $3$ & $16$ & $2$ & $2.26\pm0.20$ & $0.28\pm0.01$ & Blue diamonds\\
 $4$ & $12$ & $1.5$ & $4.38\pm0.30$ & $0.66\pm0.01$ & Green pentagrams\\
\end{tabular}
\caption{Parameters \DEL{and measured quantities }for the four series of experiments with the flattop hill cross section. \ADD{The inertial coefficient $C^{\mu}$ and the viscous damping rate $\lambda_{\textrm{h}}$ are measured in the homogeneous case.} The symbols mentioned in the last column are used in figures~\ref{fig:flattop_added_mass} and~\ref{fig:flattop_damping}. }
\label{tabular:flattop_parameters}
\end{center}
\end{table}

The \DEL{added mass}\ADD{inertial coefficient} and viscous damping \DEL{coefficients}\ADD{rate} have been measured first in a homogeneous fluid using the impulse response function analysis. As expected, the \DEL{added mass}\ADD{inertial coefficient} has been found constant for all frequencies and the viscous damping \DEL{coefficient}\ADD{rate} depends on the square root of the frequency. The results for the different fluid depths are shown in Table~\ref{tabular:flattop_parameters}.

\begin{figure}
\begin{center}
\includegraphics[width=0.5\linewidth,clip=]{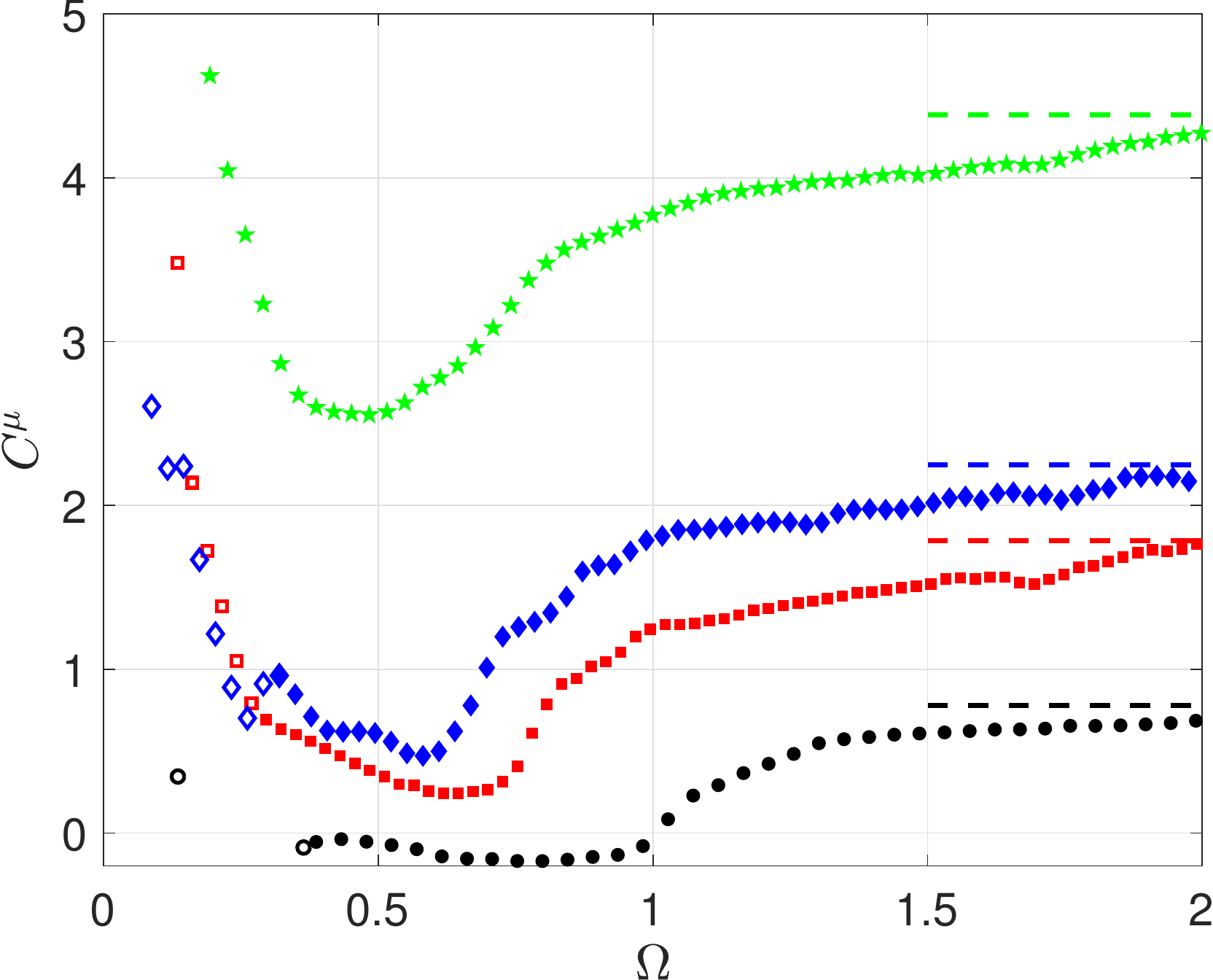}
    \caption{\ADD{Stratified case:} \DEL{Added mass}\ADD{inertial} coefficient $C^{\mu}$ as a function of the dimensionless frequency $\Omega$ for the four different series of experiments at different depths. The symbols are indicated in Table~\ref{tabular:flattop_parameters}. The frequency range where wave reflections are important corresponds to the range where the symbols are empty. The horizontal dashed lines correspond to the \DEL{added mass}\ADD{inertial coefficient} measured in a homogeneous fluid for the different depths. The colors are the same than the ones used for the symbols.\label{fig:flattop_added_mass}     }
   \end{center}
\end{figure}

Figure~\ref{fig:flattop_added_mass} shows the \DEL{added mass}\ADD{inertial} coefficients \ADD{$C^{\mu}$}, for the four series of experiments. The different symbols are indicated in Table~\ref{tabular:flattop_parameters}. As for the cylinder with a circular cross section \ADD{and the vertical flat plate} (see section\ADD{s}~\ref{disk} \ADD{and~\ref{vertical_plate}}), the \DEL{added mass}\ADD{inertial coefficient} at smaller fluid depth is larger. Moreover, for large frequency, the \DEL{added mass}\ADD{inertial} coefficients reach the asymptotic values found in the homogeneous fluid. These values are represented by horizontal dashed lines in figure~\ref{fig:flattop_added_mass}.

\begin{figure}
\begin{center}
\includegraphics[width=1\linewidth,clip=]{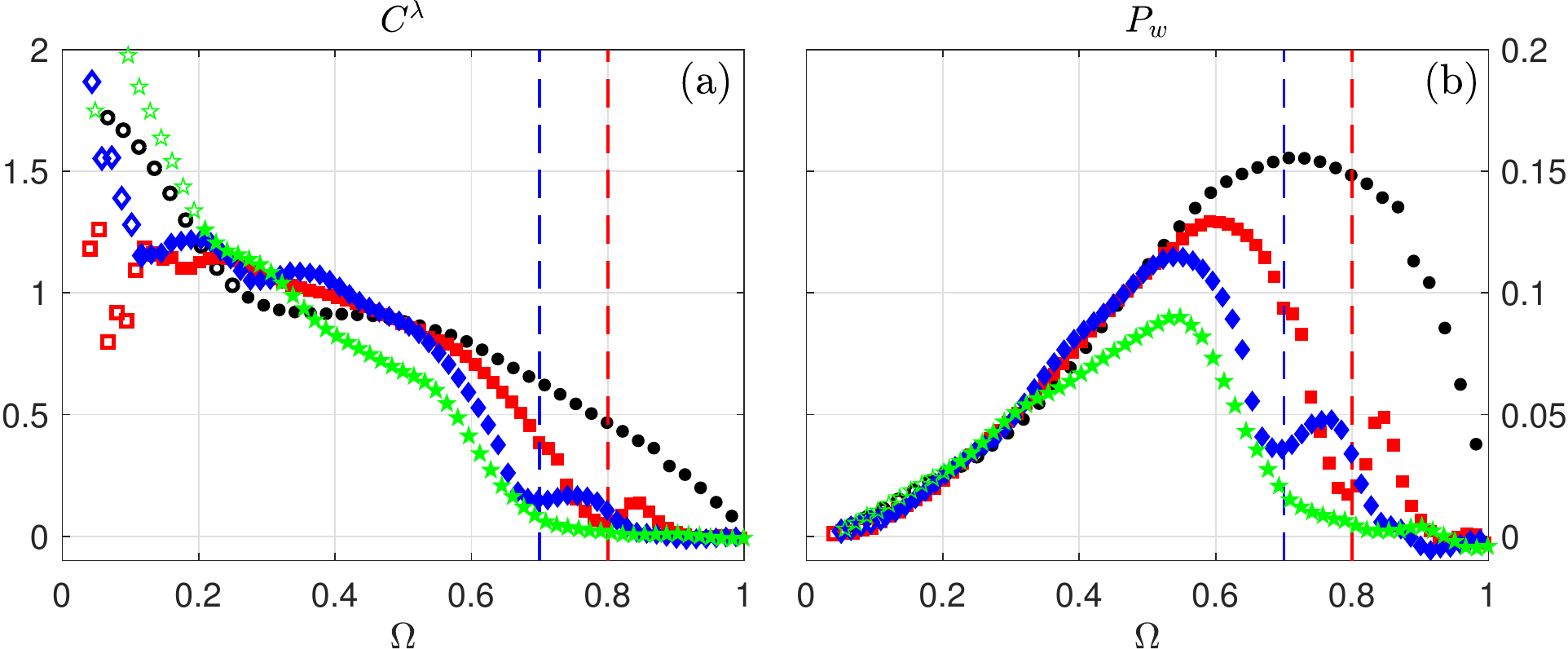}
    \caption{\ADD{Stratified case:} \ADD{wave} damping coefficient $C^{\lambda}$ (a) and radiated wave power $P_{\textrm{w}}$ (b) as a function of the dimensionless frequency $\Omega$ for the four different series of experiments at different depths. The symbols are indicated in Table~\ref{tabular:flattop_parameters}. The frequency range where wave reflections are important corresponds to the range where the symbols are empty. The two vertical dashed lines show the prediction for $\Omega_{\ell}$ for the two series of experiments at intermediate depths, using figure~\ref{fig:flattop_prediction}. 
    	\label{fig:flattop_damping}}
   \end{center}
\end{figure}

Figure~\ref{fig:flattop_damping} shows the wave damping and radiated wave power coefficients, for the four series of experiments. The different symbols are indicated in Table~\ref{tabular:flattop_parameters}. 
One can see a significant local minimum in these coefficients, at $\Omega=0.8$ for $H=20.3$~cm (red squares) and at $\Omega=0.7$ for $H=16$~cm (blue diamonds), while it is not the case for $H=95$~cm (black circles) and $H=12$~cm (green pentagrams).  This is fully consistent with the theoretical prediction, given in figure~\ref{fig:flattop_prediction}. No minimum is seen for the infinite depth experiment, because the predicted frequency lacking of tidal conversion $\Omega_{\ell}$ is too close to~$1$.
 Moreover, for the supercritical topography ($\Omega<\Omega_s$), there is no local minimum of tidal conversion. 
 Note that the wave reflection at the end of the tank affects  the signal only for $\Omega$ smaller than $0.3$ 
so that the data in the frequency range of interest is fully reliable. Since the data corresponding $\Omega<0.3$ are not reliable, we cannot make any conclusion concerning the enhancement of tidal conversion at $\Omega\rightarrow 0$ for this type of topography.

\section{Conclusions}

In this paper, we examined the effects of finite depth on tidal conversion using the concept of added mass and \DEL{the added mass and damping coefficients of four}\ADD{three}
different bodies oscillating in a uniformly stratified fluid\ADD{.}\DEL{ have been measured. }

First, we \ADD{validate our set-up using}\DEL{ consider} the case of a square-shaped cylinder with a vertically oriented diagonal in a stratified fluid of infinite depth. In the context of baroclinic tidal conversion, this case mimics an isolated bottom topography with a triangular cross-section. We correct a small error present in calculations of~\cite{ErmanyukGavrilov2002b}, and perform a series of experiments with a larger cylinder than the one used in~\cite{ErmanyukGavrilov2002b}, \ADD{observing reasonable agreement between the results.} 
\DEL{For $\Omega>1$, when no wave is emitted, the theoretical prediction is well satisfied. For $\Omega<1$, the cylinder generates internal waves which partially return back to it after reflections at the ends of the test tank. This effect corrupts the data for the frequency range $\Omega<0.4$. However, in the frequency domain of interest $0.4<\Omega<1$, we clearly observe the sub/super critical transition around $\Omega=\sqrt{2}/2$ in full agreement with the theoretical prediction. The transition is marked by a sudden drop of the radiated wave power when the topography becomes subcritical.}

Second, we investigate the effect of limited depth on the force coefficient and radiated wave power for a circular cylinder and a vertical flat plate. This study is motivated by the possibility of enhancement ~\citep{Petrelisetal2006,LlewellynSmithYoung2003} or reduction ~\citep{LlewellynSmithYoung2002} of tidal conversion for super- and subcritical topographies. In this context, the interest of circular geometry is that it always has a sub- and supercritical parts at any frequency $\Omega<1$, while the vertical flat plate corresponds to the ultimate case of supercritical topography. For the circular cylinder we have extended the previous data set~\cite{ErmanyukGavrilov2002b} to low values of depth-to-diameter ratio $H/b=1.5$ and~$1.2$. In a narrow range of frequency at $\Omega\rightarrow 0$, we observe a weak enhancement of tidal conversion: for $H/b=1.2$ it increases by roughly $15\%$ compared to the value in the fluid of infinite extent. However, the main trend observed for $0.3<\Omega<1$ exhibits the reduction of tidal conversion compared to the case of infinite fluid. For the vertical plate, we re-derive the expression for the enhancement factor due to limited depth and find a simple formula which is in full agreement with an integral expression from~\citep{LlewellynSmithYoung2003}. However, our measurements of the \ADD{wave} damping coefficient and radiated wave power performed at $H/b=2.2$ and $1.3$ have not demonstrated any convincing evidence of enhancement of tidal conversion although the expected effect far exceeds the uncertainty of measurements. On the contrary, the enhancement is present for the \DEL{added mass}\ADD{inertial coefficient} of the oscillating flat plate, especially at large oscillation frequency dynamically corresponding to unstratified fluid. However, as the oscillation frequency decreases the \DEL{added mass}\ADD{inertial coefficient} exceeds the theoretical prediction, most notably at $0<\Omega<1$ where it is expected to be identically zero regardless to the value of $H/b$. The absence of the enhancement of tidal conversion at the laboratory scale can be attributed to the final width of wave beams, which can overlap and interfere destructively in contrast to infinitely thin beams assumed in ideal-fluid theory ~\citep{LlewellynSmithYoung2003}.

Finally, we measured for the first time the force coefficients and the radiated wave power of an object with a specific cross section, inspired by~\cite{Maas2011}. We show that it exhibits a lack of tidal conversion for a specific frequency $\Omega_{\ell}$, as expected. This frequency depends on the depth of the fluid and the results are consistent with the theoretical prediction. Below a certain depth-to-height ratio $H/b$ a local minimum of the radiated wave power cannot be observed. The findings presented in this article can have some \ADD{important}\DEL{strong} consequences in the oceanographic context due to a large variety of realistic sub- and supercritical bottom topographies, including those lacking tidal conversion~\citep{Maas2011}.

\begin{acknowledgments}
{\bf Acknowledgments}

EVE gratefully acknowledges his appointment as a Marie Curie incoming fellow at Laboratoire de physique. 
This work was supported by the LABEX iMUST (ANR-10-LABX-0064) of Université de Lyon, within 
the program "Investissements d'Avenir" (ANR-11-IDEX-0007) operated by the French National 
Research Agency (ANR). This work has 
achieved thanks to the resources of PSMN from ENS de Lyon.
We thank L. Maas and H. Scolan for helpful discussions.

\end{acknowledgments}

\appendix

\section{Square cylinder\label{square}}

\ADD{This appendix presents the experiments performed in a fluid of large depth with a cylinder having a square cross section. Such experiments have already been performed by~\cite{ErmanyukGavrilov2002b}. Below we cross-compare the two data sets to validate the methodology of the present work and to correct an error present in~\cite{ErmanyukGavrilov2002b}.}

The sides of the square used in the present paper are $10$~cm long. The cylinder is fixed to the pendulum in order to have the diagonals of the square vertical or horizontal, leading to $a=b\approx14.4$~cm. The fluid depth $H$ is equal to $95$~cm. Thus, we are in a good approximation in a fluid of infinite depth, with $H/b\approx6.6$.
Experiments~\citep{ErmanyukGavrilov2002b} have been performed with a smaller cylinder, at $a=b=5.2$~cm and $H=36$~cm. The characteristic parameters of the cylinder used in the present work are given in Table~\ref{tabular:shape_parameters}\ADD{, located in the main body of the paper}. 

For a diamond-shaped cylinder the added mass in a homogeneous fluid of infinite depth is~\citep{Korotkin2010}
\begin{equation}
F_{\ast}(p)=\frac{\Gamma\left(1.5- \arctan(1/p)/\pi\right)\Gamma\left(\arctan(1/p)/\pi\right)}{\Gamma\left(0.5+\arctan(1/p)/\pi\right)\Gamma\left(1- \arctan(1/p)/\pi\right)}-1,\label{eq:square_homogeneous}
\end{equation}
where $\Gamma$ is the Euler function and $p=b/a$, i.e. the ratio between the vertical and the horizontal diagonals of the diamond. \ADD{The Euler function is defined for any complex number $z$ with a positive part as $\Gamma(z)=\int_0^{\infty}x^{z-1}\exp(-x)\textrm{d}x$.}
For the square cross-section in our set-up, $p=1$ which leads to 
$F_{\ast}(p=1)\approx1.19$ and $C^{\mu}\approx0.758$. Plugging~(\ref{eq:square_homogeneous}) into equations~(\ref{rulell}) and~(\ref{rulhyp}) and using the complex Euler function for $\Omega<1$, one can compute the solution for the square-shaped cylinder in a uniformly stratified fluid of infinite extent. For brevity, we discuss only the results for wave damping \ADD{$C^{\lambda}$ and radiated wave} power \ADD{$P_{\textrm{w}}$} (see figure~\ref{fig:lambda_eau_strat_grand_carre}), which are the quantities directly relevant to tidal conversion. 
It is worthwhile to note that, due to an unnoticed bug in the software, which returned different quantities for $\sqrt{-1}$ and $(-1)^{1/2}$, an error has been made by~\cite{ErmanyukGavrilov2002b} in calculations at $\Omega<1$. In the present paper this error is corrected.

\ADD{For $\Omega>1$, the wave damping $C^{\lambda}$ and radiated power $P_{\textrm{w}}$ coefficients are \ADD{expected to be} identically null due to absence of wave emission.} For $\Omega<1$, the solution predicts a sudden drop of the \ADD{wave} damping coefficient and the non-dimensional wave power  at the frequency corresponding to the transition from supercritical to subcritical case, at $\Omega=\sqrt{2}/2$\DEL{ (see figure~\ref{fig:lambda_eau_strat_grand_carre})}. The limit $C^{\lambda}\rightarrow 1$ at $\Omega\rightarrow 0$ is the same as for a circular cylinder \ADD{or a vertical flat plate (see sections~\ref{disk} and~\ref{vertical_plate})}, asserting that in the fluid of infinite extent this quantity is defined by the height of the obstacle and does not depend on geometry.

\begin{figure}
	\begin{center}
\includegraphics[width=0.95\linewidth,clip=]{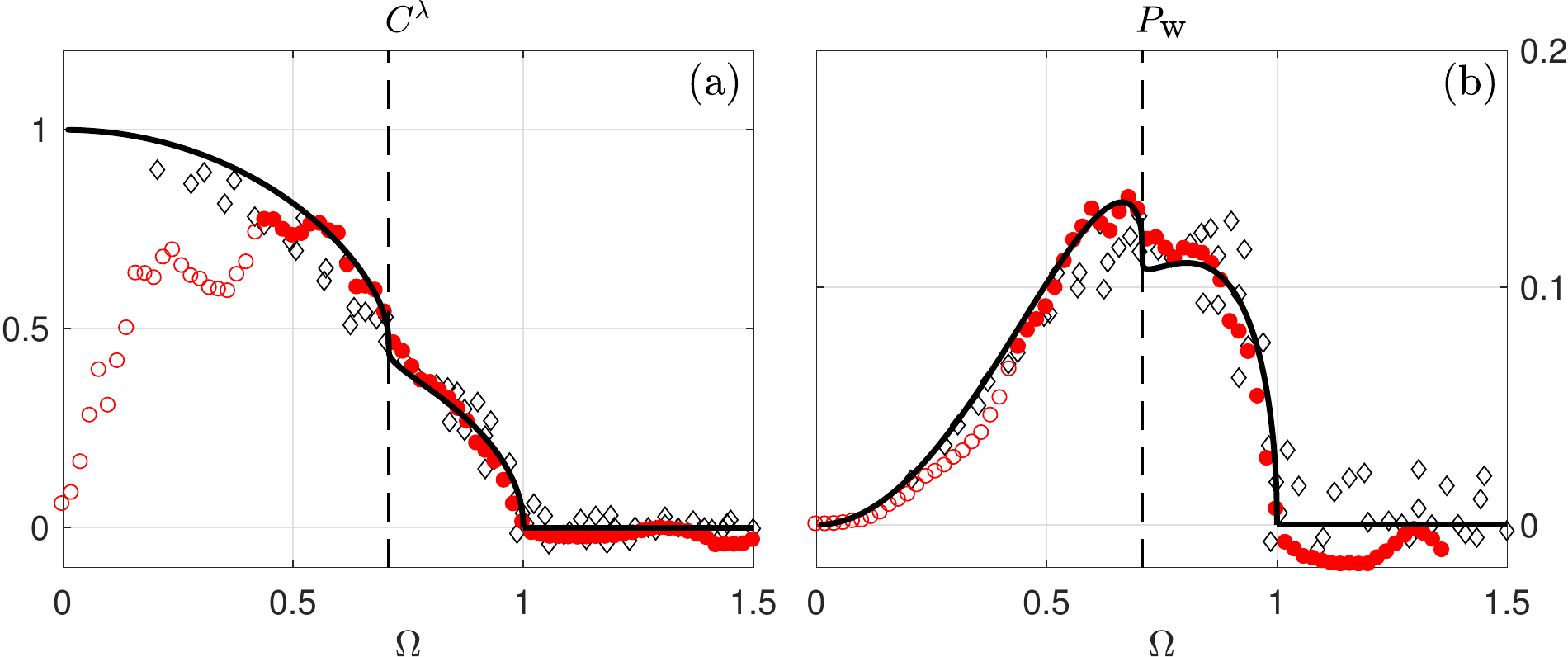}		
		\caption{\ADD{Stratified case:} \ADD{wave} damping \ADD{$C^{\lambda}$} (a) and radiated \ADD{wave} power \ADD{$P_{\textrm{w}}$} (b) \DEL{dimensionless }coefficients as a function of the dimensionless frequency $\Omega$, for the cylinder with a square cross section. The black diamonds are from~\cite{ErmanyukGavrilov2002b} while the red circles have been obtained with our experiments. The frequency range where wave reflections are important corresponds to the range where the red circles are empty. The theoretical prediction
			~(\ref{eq:square_homogeneous}) is plotted as a black line, while the vertical dashed line shows the critical frequency $\Omega=\sqrt{2}/2$. 			\label{fig:lambda_eau_strat_grand_carre}}
	\end{center}
\end{figure}

The black empty diamonds in figure~\ref{fig:lambda_eau_strat_grand_carre} are extracted from~\cite{ErmanyukGavrilov2002b} while the red circles are points obtained with the present experimental set-up. For  $\Omega>1$, the \ADD{wave} damping coefficient and the radiated power are close to zero within the experimental accuracy. The overall agreement between~\cite{ErmanyukGavrilov2002b} and the present results at $\Omega<1$ is reasonably good. Both results follow the theoretical prediction in the range $0.4<\Omega<1$.  The new set of data visualizes the singularity at $\Omega=\sqrt{2}/2\approx 0.7$\DEL{. F} \ADD{while f}or the data set from~\cite{ErmanyukGavrilov2002b}, the singularity is smoothed out. We believe that this effect occurs because of a higher relative boundary layer thickness at a smaller sylinder, acting as a "coating" for the sharp-angled shape of the cross-section. Indeed, the size of the cylinder in ~\cite{ErmanyukGavrilov2002b} is $2.7$ times smaller than in the present experiments. For the new set of data the \ADD{wave} damping coefficient exhibits increasing departure from the theoretical prediction as $\Omega$ falls below $0.4$ (see data marked by empty red circles in figure~\ref{fig:lambda_eau_strat_grand_carre}(a)). This is related to wave reflections from the ends of the test tank, which are more persistent at lower $\Omega$ and larger scale of the oscillating object.

\end{document}